\documentclass[12pt]{article}
\usepackage[english]{babel}
\usepackage{amssymb,slashed,latexsym,amsmath,multirow,color}
\usepackage{slashed}
\pdfoutput=1
\usepackage[font=footnotesize,labelsep=newline,labelfont=sc,justification=centering,position=top]{caption}

\numberwithin{equation}{section}

\newcommand{\D}{\mathcal{D}}
\newcommand{\opsi}{\bar{\psi}}

\newcommand{\Linv}{L^{-1}}
\usepackage{hyperref}
\newcommand\T{\rule{0pt}{2.6ex}}
\newcommand\B{\rule[-1.2ex]{0pt}{0pt}}
\newcommand{\cn}{{\cal N}}  
\newcommand{\ft}[2]{{\textstyle\frac{#1}{#2}}}
\def\rmi{{\rm i}}
\newsavebox{\uuunit}
\sbox{\uuunit}
    {\setlength{\unitlength}{0.825em}
     \begin{picture}(0.6,0.7)
        \thinlines
        \put(0,0){\line(1,0){0.5}}
        \put(0.15,0){\line(0,1){0.7}}
        \put(0.35,0){\line(0,1){0.8}}
       \multiput(0.3,0.8)(-0.04,-0.02){12}{\rule{0.5pt}{0.5pt}}
     \end {picture}}

\newcommand{\SO}{\mathop{\rm SO}}
\newcommand{\SU}{\mathop{\rm SU}}
\newcommand{\OSp}{\mathop{\rm OSp}}

\newcommand{\F}{\hat{F}}

\begin{document}
\begin{titlepage}
\begin{flushright}
arXiv:1101.2403
\end{flushright}
\vspace{.5cm}
\begin{center}
\baselineskip=16pt {\LARGE    Off-shell $\cn=(1,0)$, $D=6$ supergravity
from superconformal methods
}\\
\vfill
{\large Frederik Coomans$^{\dag}$, Antoine Van Proeyen$^{\dag}$
  } \\
\vfill
{\small Instituut voor Theoretische Fysica, Katholieke Universiteit Leuven,\\
       Celestijnenlaan 200D B-3001 Leuven, Belgium.
\\ \vspace{6pt}
 }
\end{center}
\vfill
\begin{center}
{\bf Abstract}
\end{center}
{\small We use the superconformal method to construct the full off-shell
action of $\cn=(1,0)$, $D=6$ supergravity, which has apart from the
graviton and the gravitino, a 2-form gauge field, a dilaton and a
symplectic Majorana spinor. We give detailed formula for superconformal
expressions that can be useful for extensions of the theory to more
matter multiplets or gauged supergravity.}
 \vfill

\hrule width 3.cm \vspace{2mm}{\footnotesize \noindent $^{\dag}$e-mails:
\{Frederik.Coomans, Antoine.VanProeyen\}@fys.kuleuven.be}
\end{titlepage}
\addtocounter{page}{1}
 \tableofcontents{}
\newpage

\section{Introduction}

In \cite{Salam:1984cj} Salam and Sezgin argued that $\cn=(1,0)$, $D=6$
Poincar{\'e} supergravity coupled to a $\rm{U}(1)$ vector multiplet
spontaneously compactifies on $\mathcal{M}_4 \times S^2$ giving rise to
chiral fermions and breaking supersymmetry down to $\cn=1$. In view of
extending this model and to obtain a more realistic compactification
scheme, general couplings of $\cn=(1,0)$, $D=6$ supergravity to different
types of matter multiplets were constructed in \cite{Bergshoeff:1985mz}.
These couplings were constructed by using the method of superconformal
tensor calculus. This method, developed in \cite{Kaku:1977pa,
Kaku:1978ea, deWit:1979ug, deWit:1980tn, deWit:1984pk, deWit:1984px},
provides a very convenient framework to study general matter couplings to
supergravity theories. After writing down the superconformal theory,
suitable gauge conditions (breaking superconformal symmetries down to
super-Poincar{\'e} symmetries) give rise to different formulations of
Poincar{\'e} supergravity.

The minimal field content of $\cn=(1,0)$, $D=6$ supergravity contains an
antisymmetric tensor with a self-dual field strength, often called chiral
2-forms. This leads to difficulties for a Lagrangian formulation, similar
to those for type IIB supergravity. We refer to \cite{Pasti:2009xc} for a
discussion on the possibilities for the Lagrangians of theories with
chiral 2-forms. The action in \cite{Bergshoeff:1985mz} is constructed by
combining the anti-self dual field strength with a self-dual field
strength that is present in a tensor multiplet. This will lead to
supergravity with a physical (non-chiral) 2-form, avoiding the chirality
problems. We will denote this theory as `off-shell $\cn=(1,0)$, $D=6$
supergravity'.

In \cite{Bergshoeff:1985mz} only the bosonic part of this action was
explicitly constructed. In this article we will discuss this method in
detail and construct the full (fermionic plus bosonic) action for minimal
6 dimensional supergravity with auxiliary fields using superconformal
tensor calculus.

The off-shell Poincar{\'e} action has several applications. When obtaining
this action by gauge fixing the redundant conformal symmetries of the
superconformal action, the $\rm{SU}(2)$ R-symmetry group is broken down
to $\rm{U}(1)$. A vector multiplet action can then be added to the
off-shell Poincar{\'e} action, gauging this $\rm{U}(1)$ R-symmetry. Hence,
one obtains an off-shell theory that is dual to the Salam-Sezgin model,
which is 6-dimensional gauged Maxwell-Einstein supergravity.

In a next step, curvature squared terms ($R^2$-terms) can be added to
this dual Salam-Sezgin model, and the influence of these terms to
solutions of the model (like the $\mathcal{M}_4 \times S^2$ solution or
brane solutions) can be studied. A convenient trick to construct
supersymmetric $R^2$-terms was developed in \cite{Bergshoeff:1986vy,
Bergshoeff:1986wc}. This $R^2$-action is off-shell, hence justifying the
need for a full off-shell formulation of Poincar{\'e} supergravity.

Another application is related to \cite{Lu:2010ct}. In this article a
three-parameter family of massive $\cn=1$ supergravities in $D=3$ is
obtained from the $S^3$ reduction of an off-shell $D=6$ Poincar{\'e}
supergravity that includes an $R^2$-term. Only the bosonic terms of
these actions are obtained via compactification. The fermionic part is
constructed afterwards via the Noether method. From a full off-shell
Poincar{\'e} action in six dimensions, also the fermionic terms of these
three dimensional theories can be reached directly from
compactification.

As said above the aim of this article is to construct an off-shell
action for Poincar{\'e} supergravity using superconformal methods. The fact
that the minimal field content of $\cn=(1,0)$, $D=6$ supergravity
contains an anti-self dual field strength is reflected in the
superconformal theory by the fact that the Standard Weyl multiplet that
is known from $D=4$ \cite{deWit:1979ug}, and $D=5$
\cite{Kugo:2000hn,Bergshoeff:2001hc,Fujita:2001kv} contains an anti-self
dual tensor $T^-_{abc}$. We denote this here as the Weyl 1 multiplet.
Combining this multiplet with a superconformal tensor multiplet, which
has a self-dual tensor  leads to a superconformal multiplet with a
physical (non-chiral) tensor (Weyl 2 multiplet). To obtain meaningful
superconformal actions, one needs also a further `compensating
multiplet'. Two choices have been discussed in \cite{Bergshoeff:1985mz}:
a hypermultiplet and a linear multiplet. In order to get an off-shell
Poincar{\'e} theory, the linear multiplet is required. The different
multiplets and couplings will be reviewed in detail in section
\ref{section_outline}.

This section also contains different tables in which the reader can
follow the set of independent fields in each step of the construction.
They contain also the counting of degrees of freedom (dof) without use of
field equations (`off-shell counting'). The final super-Poincar{\'e} theory
contains 48+48 off-shell dof. After use of the field equations they are
reduced to 16+16: graviton, scalar and 2-rank antisymmetric tensor on the
bosonic side, and gravitino and a simple spinor on the fermionic side.

In Section \ref{section_construction} the action is constructed
explicitly. First the superconformal action is built by coupling a linear
multiplet to the Weyl 1 multiplet. After gauge fixing and expressing the
action in terms of the Weyl 2 multiplet, the Poincar{\'e} action is obtained.

In Section \ref{ss:conclusions} we write down our conclusions. Appendix
\ref{ss:conventions} sets out our notation and conventions and appendix
\ref{ss:6dsca} summarizes the commutation relations of the $\cn=(1,0)$,
$D=6$ superconformal algebra. Finally, in Appendix
\ref{app:vectorfromlinear} we discuss the construction of a vector
multiplet from the components of a linear multiplet, which is used to
construct an action for the linear multiplet in Section
\ref{section_construction}.

%

\section{Outline of the procedure} \label{section_outline}

\subsection{The \texorpdfstring{$\cn=(1,0)$, $D=6$}{N=(1,0), D=6} Weyl multiplet}

In this section we will construct the $\cn=(1,0)$, $D=6$ Weyl multiplet. We
will mainly follow the outline of \cite{Bergshoeff:1985mz}. We begin our
discussion with the superconformal algebra in 6 dimensions:
$\rm{OSp}(6,2|1)$. The bosonic generators of this algebra are the generators
of the conformal group (translations $P_a$, rotations $M_{ab}$, dilatations
$D$ and special conformal transformations $K_a$) plus an $\rm{SU}(2)$ triplet
generator $U_i{}^j$. The fermionic generators are the supersymmetries
$Q^i_{\alpha}$ and the `special' supersymmetries $S^i_{\alpha}$. The nonzero
commutators between these generators and their properties are given in
appendix \ref{ss:6dsca}.

To each generator $T_A$ of the superconformal algebra we assign a gauge
field $h_{\mu}^A$ in the following way:
\begin{equation}
h_{\mu}^AT_A=e_{\mu}{}^aP_a+\ft12\omega_{\mu}{}^{ab}M_{ab}+b_{\mu}D+f_{\mu}{}^aK_a+\bar{\psi}_{\mu}^iQ_i+\bar{\phi}_{\mu}^iS_i
+{\cal V}_{\mu i}{}^jU_j{}^i\,, \label{defgaugefields}
\end{equation}
with $\psi_{\mu}^i$ and $\phi_{\mu}^i$ $\rm{SU}(2)$ Majorana-Weyl spinors
of positive and negative chirality, respectively. Using the structure
constants $f_{AB}{}^C$ of the superconformal algebra (given in appendix
B) and the basic rules
\begin{eqnarray}
\delta h_{\mu}^A &=& \partial_{\mu}\epsilon^A+\epsilon^C h_{\mu}^B f_{BC}{}^A\,, \nonumber\\
R_{\mu\nu}{}^A &=& 2\partial_{[\mu}h_{\nu]}{}^A+h_{\nu}^Ch_{\mu}^Bf_{BC}^A\,,
\end{eqnarray}
one can easily write down the (linear) transformation rules and the
curvatures $R_{\mu\nu}{}^A$ of the superconformal gauge fields given in
(\ref{defgaugefields}). Also, in order to achieve maximal irreducibility
of the superconformal gauge field configuration one imposes a maximal set
of conventional constraints \cite{Gates:1979jv, Gates:1979wg} on these
curvatures\footnote{The full expression of the linear transformation
rules and curvatures and a more detailed discussion on the conventional
constraints is given in \cite{Bergshoeff:1985mz}.}. These constraints
determine the (dependent) gauge fields $\omega_{\mu}{}^{ab}$,
$\phi_{\mu}^i$ and $f_{\mu}{}^a$ in function of the (independent) gauge
fields $e_{\mu}{}^a$, $\psi_{\mu}{}^i$, $b_{\mu}$, ${\cal
V}_{\mu}{}^{ij}$.

Counting the number of bosonic and fermionic degrees of freedom of these
gauge fields, one finds that they do not match. Hence, the algebra does
not close. Additional matter fields $T^{-}_{abc}$, $\chi^i$ and $D$ must
be added to the gauge fields in order to obtain a closed multiplet
\cite{deWit:1979ug, deWit:1980tn}. $T^{-}_{abc}$ is an antisymmetric
tensor of negative duality, $\chi^i$ is an $\rm{SU}(2)$ Majorana-Weyl
spinor of negative chirality and $D$ is a real scalar.

Starting from the linear transformation rules of the superconformal gauge
fields, the curvatures $R_{\mu\nu}{}^A$ and the matter fields
$T^{-}_{abc}$, $\chi^i$ and $D$ we can construct the full nonlinear
$\cn=(1,0)$, $D=6$ Weyl multiplet by applying an iterative procedure
outlined in \cite{Bergshoeff:1985mz}. The results are\footnote{The gauge
field and parameter of the $\SU(2)$ R-symmetry is normalized with a
factor $-2$ difference w.r.t. \cite{Bergshoeff:1985mz}.}
\begin{eqnarray}
\delta &=& \bar{\epsilon}Q+\bar{\eta}S+\lambda_DD+\ft12\lambda ^{ab}M_{ab}+\lambda_K^aK_a+\lambda_i{}^jU_j{}^i, \nonumber \\
\delta e_{\mu}{}^a &=& \frac{1}{2}\bar{\epsilon}\gamma^a\psi_{\mu}-\lambda_De_{\mu}{}^a-\lambda ^{ab}e_{\mu}{}_b, \nonumber \\
\delta \psi_{\mu}^i &=& \partial_{\mu}\epsilon^i+\frac{1}{2}b_{\mu}\epsilon^i+\frac{1}{4}\omega_{\mu}{}^{ab}\gamma_{ab}\epsilon^i+{\cal V}_{\mu}{}^i{}_j\epsilon^j +\frac{1}{24}\gamma\cdot T^-\gamma_{\mu}\epsilon^i\nonumber \\
&&+\gamma_{\mu}\eta^i-\frac{1}{2}\lambda_D\psi_{\mu}^i-\lambda^i{}_j\psi_{\mu}^j-\frac{1}{4}\lambda ^{ab}\gamma_{ab}\psi_{\mu}^i, \nonumber \\
\delta b_{\mu} &=& \partial_{\mu}\lambda_D-\frac{1}{2}\bar{\epsilon}\phi_{\mu}-\frac{1}{24}\bar{\epsilon}\gamma_{\mu}\chi+\frac{1}{2}\bar{\eta}\psi_{\mu}-2\lambda_K{}_{\mu}, \nonumber \\
\delta {\cal V}_{\mu}^{ij} &=& \partial_{\mu}\lambda^{ij}-2\lambda^{(i}{}_k{\cal V}_{\mu}^{j)k}+2\bar{\epsilon}^{(i}\phi_{\mu}^{j)}
+2\bar{\eta}^{(i}\psi_{\mu}^{j)}+\frac{1}{6}\bar{\epsilon}^{(i}\gamma_{\mu}\chi^{j)},
\end{eqnarray}
for the independent gauge fields of the multiplet. The transformation rules of the matter fields are
\begin{eqnarray}
\delta T_{abc}^-&=&3\lambda ^d{}_{[a}T_{bc]d}^-+\lambda_DT_{abc}^--\frac{1}{32}\bar{\epsilon}\gamma^{de}\gamma_{abc}\hat{R}_{de}(Q)-\frac{7}{96}\bar{\epsilon}\gamma_{abc}\chi, \nonumber \\
\delta \chi^i&=&-\frac{1}{4}\lambda ^{ab}\gamma_{ab}\chi^i-\lambda^i{}_j\chi^j+\frac{3}{2}\lambda_D\chi^i+\frac{1}{8}\bigl(\D_\mu  \gamma\cdot T^-\bigr)\gamma^{\mu}\epsilon^i \nonumber \\
&&-\frac{3}{8}\gamma\cdot\hat{R}^{ij}({\cal V})\epsilon_j+\frac{1}{4}D\epsilon^i+\frac{1}{2}\gamma\cdot T^-\eta^i\,, \nonumber \\
\delta D &=& 2\lambda_DD+\bar{\epsilon}\gamma^{\mu}\D_{\mu}\chi-2\bar{\eta}\chi\,,
\end{eqnarray}
where
\begin{eqnarray*}
\mathcal{D}_{\mu}T^-_{abc}&=&\partial_{\mu}T^-_{abc}-3\omega_{\mu}{}^d{}_{[a}T^-_{bc]d}-b_{\mu}T^-_{abc}+\frac{1}{32}\opsi_{\mu}\gamma^{de}\gamma_{abc}\hat{R}_{de}(Q)+\frac{7}{96}\opsi_{\mu}\gamma_{abc}\chi\,,  \\
\mathcal{D}_{\mu}\chi^i&=&\Bigl(\partial_{\mu}-\frac{3}{2}b_{\mu}+\frac{1}{4}\omega_{\mu}{}^{ab}\gamma_{ab}\Bigr)\chi^i+{\cal V}_{\mu}{}^i{}_j\chi^j-\frac{1}{8}\Bigl(\mathcal{D}_{\nu}\gamma\cdot T^-\Bigr)\gamma^{\nu}\psi_{\mu}^i \nonumber \\
&&+\frac{3}{8}\gamma\cdot \hat{R}^{ij}({\cal V})\psi_{\mu j}-\frac{1}{4}D\psi_{\mu}^i-\frac{1}{2}\gamma\cdot T^- \phi_{\mu}^i\,.
\end{eqnarray*}
The relevant modified curvatures $\hat{R}_{\mu\nu}{}^A$ have the form
\begin{eqnarray}
\hat{R}_{\mu\nu}{}^a(P) &=& 2\partial_{[\mu}e_{\nu]}{}^a+2b_{[\mu}e_{\nu]}{}^a+2\omega_{[\mu}{}^{ab}e_{\nu]b}-\frac{1}{2}\bar{\psi}_{\mu}\gamma^a\psi_{\nu}\,,\nonumber \\
\hat{R}_{\mu\nu}{}^i(Q) &=& 2\D_{[\mu}\psi_{\nu]}^i\,,\nonumber \\
\hat{R}_{\mu\nu}{}^{ab}(M) &=& 2\partial_{[\mu}\omega_{\nu]}{}^{ab}+2\omega_{[\mu}{}^{ac}\omega_{\nu]c}{}^b-8f_{[\mu}{}^{[a}e_{\nu]}{}^{b]}+\bar{\psi}_{[\mu}\gamma^{ab}\phi_{\nu]} \nonumber \\
&&+\bar{\psi}_{[\mu}\gamma^{[a}\hat{R}_{\nu]}{}^{b]}(Q)+\frac{1}{2}\bar{\psi}_{[\mu}\gamma_{\nu]}\hat{R}^{ab}(Q)-\frac{1}{6}e_{[\mu}{}^{[a}\bar{\psi}_{\nu]}\gamma^{b]}\chi-\frac{1}{2}\bar{\psi}_{\mu}\gamma_c\psi_{\nu}T^-{}^{abc}\,,\nonumber \\
\hat{R}_{\mu\nu}(D) &=& 2\partial_{[\mu}b_{\nu]}+4f_{[\mu}{}^ae_{\nu]a}+\bar{\psi}_{[\mu}\phi_{\nu]}+\frac{1}{12}\bar{\psi}_{[\mu}\gamma_{\nu]}\chi\,,\nonumber \\
\hat{R}_{\mu\nu}{}^{ij}({\cal V}) &=& 2\partial_{[\mu}{\cal V}_{\nu]}{}^{ij}-2{\cal V}_{[\mu}{}^{k(i}{\cal V}_{\nu]}{}^{j)}{}_k
-4\bar{\psi}_{[\mu}{}^{(i}\phi_{\nu]}{}^{j)}-\frac{1}{3}\bar{\psi}_{[\mu}{}^{(i}\gamma_{\nu]}\chi^{j)}, \label{modifiedcurvatures}
\end{eqnarray}
where\footnote{We will denote the full superconformal covariant derivative
with $\D_{\mu}$ and use $\hat{D}_{\mu}$ when we explicitly removed the term
proportional to $\phi_{\mu}^i$ from the covariant derivative. The hats on the
covariant derivatives $\hat{D}_{\mu}$ express the fact that they still
contain the terms proportional to the matter fields $D$, $\chi$ and $T^-$.}
\begin{eqnarray}
\D_{[\mu}\psi_{\nu]}^i&=&\Bigl(\partial_{[\mu}+\frac{1}{2}b_{[\mu}+\frac{1}{4}\omega_{[\mu}{}^{ab}\gamma_{ab}\Bigr)\psi_{\nu]}^i
+{\cal V}_{[\mu}{}^i{}_j\psi_{\nu]}^j-\frac{1}{24}\gamma \cdot T^- \gamma_{[\nu}\psi_{\mu]}^i-\gamma_{[\nu}\phi_{\mu]}^i  \nonumber \\
&\equiv&\hat{D}_{[\mu}\psi_{\nu]}^i-\gamma_{[\nu}\phi_{\mu]}^i\,.
\end{eqnarray}
Because of the deformation of the transformation rules and curvatures by the matter fields, the conventional constraints mentioned above must also be adapted. The following set of conventional constraints is chosen:
\begin{eqnarray}
&&\hat{R}_{\mu\nu}{}^a(P)=0\,, \nonumber \\
&&\hat{R}_{\mu\nu}{}^{ab}(M)e^{\nu}{}_b-T^-_{\mu bc}T^-{}^{abc}+\frac{1}{12}e_{\mu}{}^aD=0\,, \nonumber \\
&&\gamma^{\mu}\hat{R}_{\mu\nu}{}^i(Q)=-\frac{1}{6}\gamma_{\nu}\chi^i\,. \label{conventionalconstraints}
\end{eqnarray}
These constraints determine $\omega_{\mu}{}^{ab}$, $\phi_{\mu}^i$ and $f_{\mu}{}^a$ in function of the independent gauge fields and the matter fields:
\begin{eqnarray}
\omega_{\mu}{}^{ab}&=&2e^{\nu [a}\partial_{[\mu}e_{\nu]}{}^{b]}-e^{\rho[a}e^{b]\sigma}e_{\mu}{}^{c}\partial_{\rho}e_{\sigma c}+\frac{1}{4}\bigl(2\bar{\psi}_{\mu}\gamma^{[a}\psi^{b]}+\bar{\psi}^a\gamma_{\mu}\psi^b\bigr)+2e_{\mu}{}^{[a}b^{b]}\,,  \nonumber \\
f_{\mu}{}^a&=&\frac{1}{8}\bigl(\hat{R}'_{\mu}{}^a(M)-\frac{1}{10}e_{\mu}{}^a\hat{R}'(M)\bigr)-\frac{1}{8}T_{\mu cd}^-T^{-}{}^{acd}+\frac{1}{240}e_{\mu}{}^aD\,, \nonumber \\
\phi_{\mu}^i&=&-\frac{1}{16}\bigl(\gamma^{ab}\gamma_{\mu}-\frac{3}{5}\gamma_{\mu}\gamma^{ab}\bigr)\hat{R}'_{ab}{}^i(Q)-\frac{1}{60}\gamma_{\mu}\chi^i\,, \label{constraintsolutions}
\end{eqnarray}
where $\hat{R}'_{\mu}{}^a(M) \equiv
\hat{R}'{}_{\mu\nu}{}^{ab}(M)e^{\nu}{}_b$ and $\hat{R}'(M) \equiv
\hat{R}'_{\mu}{}^a(M)e^{\mu}{}_a$. The notation $\hat{R}'(M)$ and
$\hat{R}'(Q)$ indicates that we have omitted the $f_{\mu}{}^a$ dependent
term in $\hat{R}(M)$ and the $\phi_{\mu}^i$ dependent term in
$\hat{R}(Q)$ respectively. Hence
\begin{equation}
  \hat{R}'_{\mu\nu}{}^i(Q)= 2\hat{D}_{[\mu}\psi_{\nu]}^i\,,
 \label{hatRprimeQ}
\end{equation}
and
\begin{equation}
\hat{R}'_{\mu}{}^a{}^i(Q)=e^{\nu a}\hat{R}'_{\mu\nu}{}^i(Q).
\end{equation}
Useful `traces' of the dependent fields are
\begin{eqnarray}
&&e^{a \mu}\omega_{\mu a}{}^b=e^{-1}\partial_{\mu}(e^{\mu b}e)+\frac{1}{2}\opsi^a\gamma_a\psi^b+ 5 b^b\,,\nonumber\\
&&\gamma^{\mu}\phi_{\mu}{}^i=\frac{1}{5}\gamma^{\mu\nu}\hat{D}_{\mu}\psi^i_{\nu}-\frac{1}{10}\chi^i\,, \label{usefulexpr}\\
&&8f_a{}^a=-\frac{2}{5}\Bigl(-R+5\opsi^a\phi_a+2\opsi^{\mu }\gamma^{\nu }\hat{D}_{[\nu }\psi_{\mu ]}+\frac{5}{12}\opsi_{b}\gamma^b\chi-\frac{1}{2}\opsi_b\gamma_c\psi_aT^-{}^{abc}\Bigr)+\frac{1}{5}D.
\nonumber
\end{eqnarray}
For the last expression, we expanded the $\hat{R}'(M)$ and used
\begin{equation}
R=e^{\nu}{}_be^{\mu}{}_a\bigl(2\partial_{[\mu}\omega_{\nu]}{}^{ab}+2\omega_{[\mu}{}^{ac}\omega_{\nu]}{}_c{}^b\bigr)\,.
\end{equation}

From now on we will denote the multiplet described above as the Weyl 1
multiplet. Its independent components are summarized in Table
\ref{weyl1}. The middle column indicates the number of off-shell dof
(subtracting the gauge invariances that are indicated on the right). From
the table it is clear that the Weyl 1 multiplet constitutes a 40+40
off-shell multiplet.
\begin{center}
 \begin{table}[ht]
\caption{Fields of $\cn=(1,0)$, $D=6$ conformal supergravity in the
formulation with $D$, $\chi^i$, $T^-_{abc}$ (Weyl 1 multiplet). The Weyl
weights of the fields are given, and the off-shell dof counting.
\label{weyl1} } \centering
\begin{tabular}{cccc}
\hline
\hspace{1cm} Field \T \B \hspace{1cm}& $w$ & Number of dof & \hspace{1cm} Gauge Invariance \hspace{1cm} \\ [0.1cm]
\hline
 $e_{\mu}{}^a$ &$-1$& 15 & $P_a, M_{ab}$ \\ [0.1cm]
$b_{\mu}$ &0& 0 & $K_a$ \\ [0.1cm]
${\cal V}_{\mu}{}^i{}_j$ &0& 15 & $\SU(2)$ \\ [0.1cm]
$T_{abc}^-$ &1& 10 & \\ [0.1cm]
$D$&2 & 1 & \\ [0.1cm]
 && $-1$ & $D$\\
\hline
\multicolumn{2}{c}{\textbf{bosonic dof} \T \B }&\textbf{40}& \\
\hline
$\psi_{\mu}{}^i$&$-1/2$ & 40 & $Q^i$ \\
$\chi^i$&3/2 & 8 & \\
&& $-8$ & $S^i$ \\
\hline
\multicolumn{2}{c}{\textbf{fermionic dof} \T \B }& \textbf{40} & \\
\hline
\end{tabular}
\end{table}
\end{center}

\subsection{The Weyl multiplet with the tensor gauge field} \label{subsectionWeyl2}

We cannot use the Weyl 1 multiplet, introduced in the previous section,
to write down an off-shell action for $\cn=(1,0)$, $D=6$ supergravity.
E.g. using the linear multiplet as compensator, as we will do  below, leads to an action
that lacks kinetic terms for the matter fields $D$, $\chi^i$ and
$T^-_{abc}$. Even worse is the fact that the field equation for the
scalar $D$ gives an inconsistency (we will show this explicitly in
section \ref{subsectionGaugefixing}). We can solve both problems by
following the procedure outlined in the Introduction: coupling the Weyl 1
multiplet to a tensor multiplet. The tensor multiplet consists of a real
scalar $\sigma$, an $\rm{SU}(2)$ Majorana spinor $\psi^i$ of negative
chirality and a self-dual antisymmetric tensor field $F^+_{abc}$. The
superconformal algebra only closes on these fields modulo a number of
constraints (to be discussed below). One of these constraints can be
solved as a Bianchi identity for a new antisymmetric tensor gauge field
$B_{\mu\nu}$ defined in terms of $T^-_{abc}$ and $F^+_{abc}$. The fields
$\sigma$, $\psi^i$ and $B_{\mu\nu}$ will turn out to have proper kinetic
terms and consistent field equations, hence solving the problems
mentioned above.

Let us now take a closer look at the tensor multiplet. The full
transformations under $Q$ and $S$ are \cite{Bergshoeff:1985mz}
\begin{eqnarray}
\delta \sigma&=&\bar{\epsilon}\psi\,, \nonumber \\
\delta \psi^i&=&\frac{1}{48}\gamma \cdot F^+\epsilon^i+\frac{1}{4}\slashed{\mathcal{D}}\sigma\epsilon^i-\sigma\eta^i\,, \nonumber \\
\delta F^+_{abc}&=&-\frac{1}{2}\bar{\epsilon}\slashed{\mathcal{D}}\gamma_{abc}\psi-3\bar{\eta}\gamma_{abc}\psi\,, \nonumber \\
\D_{\mu}\sigma&=&\bigr(\partial_{\mu}-2b_{\mu}\bigl)\sigma-\bar{\psi}_{\mu}\psi\,, \nonumber \\
\D_{\mu}\psi^i&=&\bigl(\partial_{\mu}-\frac{5}{2}b_{\mu}+\frac{1}{4}\omega_{\mu}{}^{ab}\gamma_{ab}\bigr)\psi^i+{\cal V}_{\mu}{}^i{}_j\psi^j-\frac{1}{48}\gamma\cdot F^+\psi_{\mu}^i-\frac{1}{4}\slashed{\mathcal{D}}\sigma\psi_{\mu}^i+\sigma\phi_{\mu}^i\,. \label{tensormultiplet}
\end{eqnarray}
The algebra only closes on the fields when the following closed set of
independent constraints is imposed
\begin{eqnarray}
&\Gamma^i \equiv \slashed{\mathcal{D}}\psi^i-\frac{1}{6}\sigma\chi^i-\frac{1}{12}\gamma\cdot T^-\psi^i=0\,,& \nonumber \\
&\mathcal{C} \equiv (\D^a\D_a-\frac{1}{6}D\bigr)\sigma+\frac{1}{3}F^+ \cdot T^-+\frac{7}{6}\bar{\chi}\psi=0\,,& \nonumber \\
&G_{ab} \equiv \D^c\bigl(F^+_{abc}-2\sigma T^-_{abc}\bigr)-\hat{\bar{R}}_{ab}(Q)\psi-\frac{1}{6}\bar{\chi}\gamma_{ab}\psi=0\,.& \label{tensorconstraints}
\end{eqnarray}
The first two constraints can be used to define $\chi^i$ and $D$ as a
function of fields of the tensor multiplet\footnote{Note that we can only
solve the first constraint in the domain $\sigma\neq 0$.}\footnote{To obtain this expression we used the expression for $f_{\mu}{}^a$ in (\ref{constraintsolutions}).}:
\begin{eqnarray}
D&=&\frac{15}{4}\sigma^{-1}\Bigl(\partial^a\D_a\sigma-3b^a\D_a\sigma+\omega_a{}^{ab}\D_b\sigma-\frac{1}{5}\sigma R+\frac{1}{3}F^+ \cdot T^- \nonumber \\
&&+\sigma\opsi^a\phi_a+\frac{2}{5}\sigma\opsi^\mu \gamma^\nu \hat{D}_{[\nu }\psi_{\mu ]}-\frac{1}{10}\sigma\opsi_b\gamma_c\psi_aT^{-abc} \nonumber \\
&&-\opsi^{\mu}\D_{\mu}\psi+\frac{1}{24}\opsi\gamma\cdot T^- \gamma^{\mu}\psi_{\mu}+\opsi\gamma^{\mu}\phi_{\mu}+\frac{7}{6}\bar{\chi}\psi\Bigr)\,, \label{tensorD}
\end{eqnarray}
and\footnote{Note that the $\slashed{\mathcal{D}}\psi^i$-term in
$\Gamma^i$ contains a term proportional to $\phi_{\mu}^i$. We replaced
$\phi_{\mu}^i$ by its solution (\ref{constraintsolutions}) to obtain the
expression for $\chi^i$. 
}
\begin{equation}
\chi^i=\frac{15}{4}\sigma^{-1}\hat{\slashed{D}}\psi^i+\frac{3}{8}\gamma^{ab}\hat{R}'{}_{ab}^i(Q)-\frac{5}{16}\sigma^{-1}\gamma \cdot T^-\psi^i\,, \label{tensorchi}
\end{equation}
where
\begin{equation}
\hat{D}_{\mu}\psi^i=\Bigl(\partial_{\mu}-\frac{5}{2}b_{\mu}+\frac{1}{4}\omega_{\mu}{}^{ab}\gamma_{ab}\Bigr)\psi^i+{\cal V}_{\mu}{}^i{}_j\psi^j-\frac{1}{48}\gamma \cdot F^+\psi_{\mu}^i-\frac{1}{4}\slashed{\mathcal{D}}\sigma\psi^i_{\mu}.
\end{equation}
This implies that the $S$-gauge field $\phi _\mu ^i$ in this Weyl
multiplet gets the value
\begin{equation}
  16\phi _\mu ^i =\left(  -\gamma ^{ab}\gamma _\mu+\frac12 \gamma _\mu\gamma ^{ab}\right) \hat{R}'{}_{ab}^i(Q)+\gamma _\mu \sigma ^{-1}\left( -
  \hat{\slashed{D}}\psi^i+\frac{1}{12} \gamma \cdot T^-\psi^i\right) \,.
 \label{phimuversion2}
\end{equation}
The following consequence is useful for the calculations below:
\begin{equation}
  16\gamma ^\mu \phi _\mu ^i =\gamma ^{ab}\hat{R}'{}_{ab}^i(Q)-6\sigma ^{-1}\hat{\slashed{D}}\psi^i+\ft12\sigma^{-1}\gamma \cdot T^-\psi^i\,.
 \label{gamphiversion2}
\end{equation}

The third
constraint (\ref{tensorconstraints}) can be solved as a Bianchi identity
\begin{equation}
F^+_{\mu\nu\rho}+2\sigma T^-_{\mu\nu\rho}=3\partial_{[\mu}B_{\nu\rho]}+3\bar{\psi}_{[\mu}\gamma_{\nu\rho]}\psi+\frac{3}{2}\bar{\psi}_{[\mu}\gamma_{\nu}\psi_{\rho]}\sigma\equiv
\hat{F}_{\mu\nu\rho}(B)\,. \label{weyl2FB}
\end{equation}
$B_{\mu\nu}$ is a newly introduced antisymmetric tensor gauge field,
which transforms as
\begin{equation}
\delta B_{\mu\nu}=-\bar{\epsilon}\gamma_{\mu\nu}\psi-\bar{\epsilon}\gamma_{[\mu}\psi_{\nu]}\sigma
+2\partial_{[\mu}\Lambda_{\nu]}\,,
\end{equation}
where $\Lambda_{\mu}$ denotes the gauge invariance of $B_{\mu\nu}$. From this we can determine $F^+_{\mu\nu\rho}$ and $T^-_{\mu\nu\rho}$ in terms of $B_{\mu\nu}$:
\begin{eqnarray}
F^+_{\mu\nu\rho}&=\frac{1}{2}\Bigl(\hat{F}_{\mu\nu\rho}(B)+\tilde{\hat{F}}_{\mu\nu\rho}(B)\Bigr)=&\hat{F}^+_{\mu\nu\rho}(B)\,,\nonumber\\
T^-_{\mu\nu\rho}&=\frac{1}{4}\sigma^{-1}\Bigl(\hat{F}_{\mu\nu\rho}(B)-\tilde{\hat{F}}_{\mu\nu\rho}(B)\Bigr)=&\ft{1}{2}\sigma^{-1}\hat{F}^-_{\mu\nu\rho}(B)\,. \label{tensorB}
\end{eqnarray}
Table \ref{tensor} summarizes the different fields and constraints of the
tensor multiplet and their respective dof.
\begin{center}
\begin{table}[ht]
\caption{Fields and constraints of the tensor multiplet. The Weyl
weight is given and the counting of off-shell and on-shell dof.
\label{tensor}} \centering
\begin{tabular}{cccc}
\hline
\hspace{1cm} Field/Constraint \T \B \hspace{1cm}&$w$ & Off-shell dof & \hspace{1cm} On-shell dof \hspace{1cm} \\ [0.1cm]
\hline
 $\sigma$&2 & 1 &  1 \\ [0.1cm]
$F^+_{abc}$ &3& 10 &  3 \\ [0.1cm]
$\mathcal{C}$ && $-1$ &  \\ [0.1cm]
$G_{ab}$& & $-10$ & \\ [0.1cm]
\hline
\multicolumn{2}{c}{\textbf{bosonic dof} \T \B} & \textbf{0} & \textbf{4} \\
\hline
$\psi^i$ &5/2& 8 & 4 \\ [0.1cm]
$\Gamma^i$ && $-8$ & \\
\hline
\multicolumn{2}{c}{\textbf{fermionic dof} \T \B} &\textbf{0}& \textbf{4} \\
\hline
\end{tabular}
\end{table}
\end{center}
In this way, we coupled the Weyl 1 multiplet to the tensor multiplet and
we obtained a different Weyl multiplet, which we will call the Weyl 2
multiplet. The latter has the same gauge fields of the superconformal
group but a different matter sector. The Weyl 2 multiplet has matter
fields $\sigma$, $\psi^i$ and $B_{\mu\nu}$. The construction of the Weyl
2 multiplet is summarized in Table \ref{Weyl2}. Each column represents
the different bosonic (upper part of the table) and fermionic (lower
part) fields of the multiplet denoted at the top of the table. The
off-shell dof (modulo the gauge transformations denoted in Table
\ref{weyl1}) are displayed between the brackets. Note that for the tensor
multiplet we also included the constraints (\ref{tensorconstraints})
because they restrict the number of dof. From the table it is clear that
the Weyl 2 multiplet is also a 40+40
off-shell multiplet. 
The dependent fields are
\begin{itemize}
  \item $\omega_{\mu}{}^{ab}$ as given in (\ref{constraintsolutions});
  \item $F^+_{\mu \nu \rho }$ and $T^-_{\mu \nu \rho }$ determined by
  (\ref{weyl2FB});
  \item $\chi ^i$ and $\phi _\mu ^i$, see  (\ref{tensorchi}) and (\ref{phimuversion2}) where $F^+_{\mu \nu \rho }$ and $T^-_{\mu \nu \rho
  }$ are the expressions of the previous item.
  \item $D$ as given in (\ref{tensorD})
  \item $f_{\mu}{}^a$, see (\ref{constraintsolutions}), where all the
  previous expressions are used.
\end{itemize}
%
%
\begin{center}
\begin{table}[ht]
\caption{Construction of the Weyl 2 multiplet from the Weyl 1 multiplet
and the tensor multiplet, the numbers in parentheses denote the off-shell
dof. The Weyl weight of the fields of the last multiplet are given.
\label{Weyl2}} \centering
\begin{tabular}{|c|c|c|c|cr|}
\hline
\hspace{1cm} Weyl 1 \T \B \hspace{1cm} && \hspace{1cm} Tensor \hspace{1cm} && \hspace{1cm} Weyl 2 \hspace{1cm}&$w$ \\ [0.1cm]
\hline
$e_{\mu}{}^a$ (15) &&  && $e_{\mu}{}^a$ (15)&$-1$ \\ [0.1cm]
$b_{\mu}$ (0) &&  && $b_{\mu}$ (0) &0\\ [0.1cm]
${\cal V}_{\mu}{}^i{}_j$ (15) &&  && ${\cal V}_{\mu}{}^i{}_j$ (15)&0 \\ [0.1cm]
$T^-_{abc}$ (10) & $+$ & $F^+_{abc}$ (10) & $\longrightarrow$ && \\ [0.1cm]
  && $G_{ab}$ ($-10$) &$\longrightarrow$&  $B_{\mu\nu}$ (10)&0 \\ [0.1cm]
$D$ (1) & $+$ & $\mathcal{C}$ ($-1$) & $\longrightarrow$ & $\ast$& \\ [0.1cm]
  && $\sigma$ (1) && $\sigma$ (1)&2 \\ [0.1cm]
dilatations ($-1$) &&  && dilatations ($-1$)& \\
\hline
\textbf{40} \T \B && \textbf{0} && \textbf{40}& \\
\hline
$\psi_{\mu}{}^i$ (40) &&  && $\psi_{\mu}{}^i$ (40) &$-1/2$ \\
$\chi^i$ (8) & $+$ & $\Gamma^i$ ($-8$) & $\longrightarrow$ & $\ast$& \\
 && $\psi^i$ (8) && $\psi^i$ (8) &5/2\\ [0.1cm]
S-susy ($-8$) &&  && S-susy ($-8$)& \\
\hline
\textbf{40} \T \B &&\textbf{0} && \textbf{40}& \\
\hline
\end{tabular}
\end{table}
\end{center}

\subsection{The linear multiplet} \label{sec:linear}

The superconformal tensor calculus procedure prescribes that, in order to
obtain an action for supergravity, we need to couple the Weyl 2 multiplet
to a compensator multiplet. Fixing the gauges of the redundant symmetries
of the superconformal group ($D$, $K_a$ and $S^i$) then amounts to fixing
a number of components of this compensator multiplet. The remaining
components will then appear as auxiliary fields in the final off-shell
formulation.

We will use a linear multiplet as compensator because this is an
off-shell multiplet. The linear multiplet consists of a triplet scalar
$L^{ij}$, an $\rm{SU}(2)$ Majorana spinor $\varphi^i$ of negative
chirality and a constrained vector $E_a$. The constraint, to be discussed
below, can be solved\footnote{This is only possible if the linear
multiplet is inert under gauge transformations. But as we do not assume
any internal gauge symmetry this poses no problem.} in terms of an
antisymmetric tensor gauge field $E_{\mu\nu\rho\sigma}$. The full $Q$ and
$S$ transformations are \cite{Bergshoeff:1985mz}
\begin{eqnarray}
\delta L^{ij}&=&\bar{\epsilon}^{(i}\varphi^{j)}, \nonumber \\
\delta \varphi^i&=&\frac{1}{2}\slashed{\mathcal{D}}L^{ij}\epsilon_j-\frac{1}{4}\gamma^aE_a\epsilon^i-4L^{ij}\eta_j\,, \nonumber \\
\delta E_a&=&\bar{\epsilon}\gamma_{ab}\D^b\varphi+\frac{1}{24}\bar{\epsilon}\gamma_a\gamma\cdot T^- \varphi-\frac{1}{3}\bar{\epsilon}^i\gamma_a\chi^jL_{ij}-5\bar{\eta}\gamma_a\varphi\,, \nonumber \\
\D_{\mu} L^{ij}&=&\bigl(\partial_{\mu}-4b_{\mu}\bigr)L^{ij}+2{\cal V}_{\mu}{}^{(i}{}_kL^{j)k}-\opsi_{\mu}{}^{(i}\varphi^{j)}\,, \nonumber \\
\D_{\mu} \varphi^i&=&\bigl(\partial_{\mu}-\frac{9}{2}b_{\mu}+\frac{1}{4}\omega_{\mu}{}^{ab}\gamma_{ab}\bigr)\varphi^i-{\cal V}_{\mu}^{ij}\varphi_j \nonumber \\
&&-\frac{1}{2}\slashed{\mathcal{D}}L^{ij}\psi_{\mu j}+\frac{1}{4}\gamma^aE_a\psi_{\mu}^i+4L^{ij}\phi_{\mu j}\,. \label{transformationslinear}
\end{eqnarray}
The algebra closes if $E_a$ satisfies following $Q$ and $S$ invariant constraint
\begin{eqnarray}
&&\D^aE_a-\frac{1}{2}\bar{\varphi}\chi=0\,, \nonumber \\
&& \D_{\mu}E_a=\bigl(\partial_{\mu}-5b_{\mu}\bigr)E_a+\omega_{\mu ab}E^b-\opsi_{\mu}\gamma_{ab}\D^b\varphi-\frac{1}{24}\opsi_{\mu}\gamma_a\gamma\cdot T^-\varphi\,,\nonumber \\
&&+\frac{1}{3}\opsi_{\mu}^i\gamma_a\chi^jL_{ij}+5\bar{\phi}_{\mu}\gamma_a\varphi\,.
\end{eqnarray}
The solution for $E_a$ in terms of $E_{\mu\nu\rho\sigma}$ is
\begin{equation}
E_a=\frac{1}{24} e^{-1}e_{\mu}{}_a\varepsilon^{\mu\nu\rho\sigma\lambda\tau}\mathcal{D}_{\nu}E_{\rho\sigma\lambda\tau},
\end{equation}
where $e$ is the square root of the metric determinant. Note that there is a gauge invariance
\begin{equation}
\delta_{\Lambda} E_{\mu\nu\rho\sigma}=4\partial_{[\mu}\Lambda_{\nu\rho\sigma]}.
\end{equation}
We will also be using the dual 2-index field
\begin{eqnarray}
E_{\mu\nu\rho\sigma}&=&\frac{1}{2} e\varepsilon_{\mu\nu\rho\sigma\lambda\tau}E^{\lambda\tau}\,, \nonumber \\
E^a&=&e_{\mu}{}^a\D_{\nu}E^{\mu\nu}\,, \nonumber \\
\delta E^{\mu\nu}&=&\bar{\epsilon}\gamma^{\mu\nu}\varphi+\opsi_{\rho}^i\gamma^{\mu\nu\rho}\epsilon^jL_{ij}+\partial_{\rho}\bigl(e^{-1}\tilde{\Lambda}^{\mu\nu\rho}\bigr)\,, \nonumber \\
\D_{\nu}E^{\mu\nu}&=&\partial_{\nu}E^{\mu\nu}-\opsi_{\nu}\gamma^{\mu\nu}\varphi-\frac{1}{2}\opsi_{\rho}^i\gamma^{\mu\nu\rho}\psi_{\nu}^jL_{ij}\,. \label{lineardualE}
\end{eqnarray}
The different components of the linear multiplet are summarized in Table
\ref{linear}. The construction of Poincar{\'e} supergravity using the linear
multiplet as compensator is summarized in Table \ref{poincare}. The two
columns on the left denote the Weyl 2 multiplet, discussed in the previous
section, coupled to the linear multiplet. The first of these two columns
displays the fields and their off-shell dof modulo the gauge transformations
which are depicted in the second column. The next two columns describe the
gauge fixing. The first of these columns contains the gauge choices leading
to a fixing of the symmetries denoted in the next column. The final three
columns of the table depict the fields of the final off-shell formulation of
Poincar{\'e} supergravity, first again with their off-shell dof (i.e. modulo the
gauge transformations denoted in the middle column) and finally their
on-shell dof. From this table it is clear that we end up with a 48+48
off-shell
description of Poincar{\'e} supergravity, describing 16+16 on-shell dof. 
\begin{center}
\begin{table}[ht]
\caption{Fields of the linear multiplet. \label{linear}}
\centering
\begin{tabular}{cccc}
\hline
\hspace{1cm} Field \T \B \hspace{1cm}&$w$ & Off-shell dof & \hspace{1cm} On-shell dof \hspace{1cm} \\ [0.1cm]
\hline
 $L^{ij}$&4 & 3 &  3 \\ [0.1cm]
$E_{\mu\nu\rho\sigma}$&0 & 5 &  1 \\ [0.1cm]
\hline
\multicolumn{2}{c}{\textbf{bosonic dof} \T \B }& \textbf{8} & \textbf{4} \\
\hline
$\varphi^i$&9/2 & 8 & 4 \\ [0.1cm]
\hline
\multicolumn{2}{c}{\textbf{fermionic dof} \T \B} &\textbf{8}& \textbf{4} \\
\hline
\end{tabular}
\end{table}
\end{center}

\begin{center}
 \begin{table}[ht]
\caption{Construction of Poincar{\'e} supergravity from the Weyl 2 multiplet
coupled to a linear compensator multiplet. The fields and symmetries of
this setup are denoted in the first two columns, the third and fourth
columns contain the gauge choices and the respective fixed symmetries,
the  last columns denote the fields and symmetries of Poincar{\'e}
supergravity, and the number of on-shell dof. \label{poincare}}
\centering
\begin{tabular}{|c|c|c|c|c|c|c|}
\hline
\multicolumn{2}{|c|}{\hspace{1cm} Weyl 2 $\times$ Linear \hspace{1cm}} \T \B & \multicolumn{2}{c}{\hspace{1cm} Gauge fixing \hspace{1cm}} & \multicolumn{3}{|c|}{\hspace{1cm} Poincar{\'e} \hspace{1cm}} \\ [0.1cm]
\hline
$e_{\mu}{}^a$ (15) & $P_a$, $M_{ab}$ & && $e_{\mu}{}^a$ (15) & $P_a$, $M_{ab}$ &9\\ [0.1cm]
$b_{\mu}$ (0) & $K_a$ & $b_{\mu}=0$ & $K_a$ & $\ast$ &&\\ [0.1cm]
${\cal V}_{\mu}{}^i{}_j$ (15) & $\rm{SU}(2)$ & && ${\cal V}_{\mu}{}^i{}_j$ (17) & $\rm{SO}(2)$ &\\ [0.1cm]
$\sigma$ (1) &&  && $\sigma$ (1) &&1\\ [0.1cm]
$B_{\mu\nu}$ (10) & $\Lambda_{\mu}$ &  && $B_{\mu\nu}$ (10) & $\Lambda_{\mu}$&6 \\ [0.1cm]
dilatations ($-1$) &&  && $\ast$ &&\\
$L^{ij}$ (3) && $L^{ij}=\frac{1}{\sqrt{2}}\delta^{ij}$ & $D$, $\rm{SU}(2)/\rm{SO}(2)$ & $\ast$ &&\\
$E_{\mu\nu\rho\sigma}$ (5) & $\tilde{\Lambda}_{\mu\nu\rho}$ &  && $E_{\mu\nu\rho\sigma}$ (5) & $\tilde{\Lambda}_{\mu\nu\rho}$& \\
\hline
\textbf{48} \T \B && && \textbf{48} &&\textbf{16}\\
\hline
$\psi_{\mu}{}^i$ (40) & $Q^i$ &  && $\psi_{\mu}{}^i$ (40)  & $Q^i$&12 \\
$\psi^i$ (8) &&  && $\psi^i$ (8) &&4\\
S-susy ($-8$) &&  && $\ast$& &\\
$\varphi^i$ (8) && $\varphi^i=0$ & $S^i$ & $\ast$ &&\\
\hline
\textbf{48} \T \B && && \textbf{48} &&\textbf{16}\\
\hline
\end{tabular}
\end{table}
\end{center}

\section{Explicit construction of the action} \label{section_construction}

In section \ref{subsectionConstructionsca} we will construct a
superconformal action for the Weyl 1 multiplet coupled to a linear
compensator. After applying the gauge fixing procedure in section
\ref{subsectionGaugefixing} we will show explicitly the problems related
to the Weyl 1 multiplet mentioned in section \ref{subsectionWeyl2}. We
then go to a formulation in terms of the Weyl 2 multiplet by using the
relations (\ref{tensorD}), (\ref{tensorchi}) and (\ref{tensorB}). This is
discussed in section \ref{subsectionPoincareWeyl2}.

\subsection{Construction of the superconformal action} \label{subsectionConstructionsca}

\subsubsection{Density formula}
We need an expression constructed from the components of the linear
multiplet that can be used as a superconformal action. In
\cite{Bergshoeff:1985mz} a density formula is given for the product of a
vector multiplet $W_{\mu}$, $\Omega^i$, $Y^{ij}$ and a linear multiplet
$L^{ij}$, $\varphi^i$, $E_a$ :
\begin{equation}
e^{-1}\mathcal{L}_{VL}=Y_{ij}L^{ij}+2\bar{\Omega}\varphi-L^{ij}\bar{\psi}_{\mu i}\gamma^{\mu}\Omega_j+\frac{1}{4}F_{\mu\nu}(W)E^{\mu\nu}. \label{densityformula}
\end{equation}
The next step is then to construct a vector multiplet from the components of the linear multiplet
\begin{eqnarray}
\Omega^i&=&\Omega^i(L^{ij}, \varphi^i, E_a)\,, \nonumber \\
Y^{ij}&=&Y^{ij}(L^{ij}, \varphi^i, E_a)\,, \nonumber \\
\hat{F}_{\mu\nu}&=&\hat{F}_{\mu\nu}(L^{ij}, \varphi^i, E_a)\,, \label{vectorfromlincomp}
\end{eqnarray}
and to plug in these components into (\ref{densityformula}) to obtain a
superconformal action for the linear multiplet. Explicit expressions for
(\ref{vectorfromlincomp}) were constructed in \cite{Bergshoeff:1985mz}
and are summarized in appendix \ref{app:vectorfromlinear} (equations
(\ref{vectorlinear})). Note that
$\hat{F}_{\mu\nu}=F_{\mu\nu}+2\opsi_{[\mu}\gamma_{\nu]}\Omega$ is the
supercovariant field strength of $W_{\mu}$. We also define
\begin{equation}
L=\bigl(L_{ij}L^{ij}\bigr)^{1/2}.
\end{equation}

\subsubsection{Bosonic part}
Using (\ref{densityformula}) and (\ref{vectorlinear}) and retaining only
the bosonic terms we obtain
\begin{eqnarray}
e^{-1}\mathcal{L}_{\rm bos} &=& -\D_a\D^aL+L^{-1}\D_aL^{ij}\D^aL_{ij}+L^{-3}L_{ij}L_{kl}\Bigl(\D^aL^{k(i}\D_aL^{j)l}-\D_aL^{ij}\D^aL^{kl}\Bigr) \nonumber \\
             && +\frac{1}{4}L^{-1}E_aE^a-\frac{1}{3}L D-L^{-3}L_{ij}E_aL^{k(i}\D^a L^{j)}{}_k  \nonumber \\
             && +\frac{1}{4}E^{\mu\nu}\Bigl(-2\partial_{\mu}L^{ij}L^k{}_j\partial_{\nu}L_{ik}L^{-3}-2\partial_{[\mu}\left(E_{\nu]}L^{-1}-2{\cal V}_{\nu]ij}L^{ij}L^{-1}\right)\Bigr)\,.
\end{eqnarray}
We also know from equation (\ref{lineardualE}) that
\begin{equation}
E^a=e_{\mu}{}^a\D_{\nu}E^{\mu\nu}=e_{\mu}{}^a\partial_{\nu}E^{\mu\nu}+\text{fermionic terms}. \label{fermionicterms}
\end{equation}
After partial integration and using the identity (\ref{LL}), we can write
the bosonic action as
\begin{eqnarray}
e^{-1}\mathcal{L}_{\rm bos} &=& -\D_a\D^aL+\frac{1}{2}L^{-1}\D_aL^{ij}\D^aL_{ij}\nonumber \\
             &&-\frac{1}{3}LD-\frac{1}{4}L^{-1}E_aE^a  +E^{\mu}{\cal V}_{\mu ij}L^{ij}L^{-1}-\frac{1}{2}E^{\mu\nu}\partial_{\mu}L^{ij}L^k_j\partial_{\nu}L_{ik}L^{-3}. \label{bosonicconformal}
\end{eqnarray}

\subsubsection{Fermionic part}
In the next section we will discuss the gauge fixing procedure. To fix the S-gauge we will choose $\varphi=0$. Hence, in order to clarify calculations, we will drop all terms directly proportional to $\varphi$ already at this stage of the calculation. To write down the fermionic part of the superconformal Lagrangian we need the fermionic terms of $F_{\mu\nu}(L)$. We obtain
\begin{eqnarray}
F_{\mu\nu}(L)&=&\text{(bosonic terms)} \nonumber \\
          &&-8L^{-1}L_{ij}\opsi_{[\mu}^i\phi_{\nu]}^j-\frac{4}{3}L^{-1}L_{ij}\opsi_{[\mu}^i\gamma_{\nu]}\chi^j+2L^{-1}\opsi_{[\mu}\gamma_{\nu]}\slashed{\mathcal{D}}\varphi-2\Linv\opsi_{[\mu}\D_{\nu]}\varphi \nonumber \\
          &&+\frac{1}{2}L^{-1}\opsi_{[\mu}\gamma^b\psi_{\nu]}E_b-2\opsi_{[\mu}\gamma_{\nu]}\Omega(L^{ij}, \varphi^i, E_a), \label{fieldstrenght}
\end{eqnarray}
where the first term in the third line is obtained by use of the solutions of the conventional constraints (\ref{constraintsolutions}). By use of (\ref{vectorlinear}) we can write
\begin{eqnarray}
-2\opsi_{[\mu}\gamma_{\nu]}\Omega(L^{ij}, \varphi^i, E_a)&=&-2L^{-1}\opsi_{[\mu}\gamma_{\nu]}\slashed{\mathcal{D}} \varphi+\frac{4}{3}L^{-1}\opsi_{[\mu}^i\gamma_{\nu]}L_{ij}\chi^j.
\end{eqnarray}
Filling in the above equation into equation (\ref{fieldstrenght}) and
using (\ref{transformationslinear}) we find that all fermionic terms of
$F_{\mu\nu}(L)$ drop. Using the density formula (\ref{densityformula}),
we obtain the fermionic part of the action which, at this stage of the
calculation, looks like (remember that we have put $\varphi$ to zero)
\begin{eqnarray}
e^{-1}\mathcal{L}_{\rm ferm}&=&-\frac{1}{4}\Linv \opsi_{\rho}^i\gamma^{\mu\nu\rho}\psi_{\nu}^j L_{ij} E_{\mu}
+\frac{1}{2}\Linv L^{ij}L_{kl}\opsi_{\rho}^k\gamma^{\mu\nu\rho}\psi_{\nu}^l {\cal V}_{\mu ij}-\frac{1}{3}L\opsi_{\mu}\gamma^{\mu}\chi \nonumber \\
             &&-\Linv L_{ij}\opsi_{\mu}^i\gamma^{\mu}\gamma^{\nu}\Bigl(-\frac{1}{2}\slashed{\mathcal{D}}L^{jk}\psi_{\nu}{}_k+\frac{1}{4}\gamma^{\rho}E_{\rho}\psi_{\nu}{}^j+4L^{jk}\phi_{\nu}{}_k\Bigr). \label{fermionicconformal1}
\end{eqnarray}
Note that the two first terms come from the step done in (\ref{fermionicterms}). In the last line we also used (\ref{transformationslinear}).

\subsubsection{
Superconformal action} \label{ssFinalform}

We will now prepare for the gauge fixing procedure by writing out the
covariant derivatives and dependent fields. The bosonic superconformal
Lagrangian in (\ref{bosonicconformal}) can be rewritten, using
(\ref{dadaL}), in a form that is most convenient for the gauge fixing
procedure
\begin{eqnarray}
e^{-1}\mathcal{L}_{\rm bos} &=& L^{-3}L_{ij}L_{kl}\D_aL^{kl}\D^aL^{ij}-\Linv \D_a L_{ij} \D^a L^{ij}-\Linv L_{ij}\D_a\D^a L^{ij} +\frac{1}{2}L^{-1}\D_aL^{ij}\D^aL_{ij}\nonumber \\
             &&-\frac{1}{3}LD-\frac{1}{4}L^{-1}E_aE^a  +E^{\mu}{\cal V}_{\mu ij}L^{ij}L^{-1}-\frac{1}{2}E^{\mu\nu}\partial_{\mu}L^{ij}L^k{}_j\partial_{\nu}L_{ik}L^{-3}\,. \label{bosonicconformal2}
\end{eqnarray}

The fermionic terms in these covariant derivatives will be collected in a
new fermionic Lagrangian, which contains not only the terms in
(\ref{fermionicconformal1}), but also terms from the `bosonic' part
(\ref{bosonicconformal2}) due to terms quadratic in fermions in covariant
derivatives or dependent bosonic fields.

We  mentioned that the choice for the S-gauge will be $\varphi=0$. We
dropped already terms proportional to $\varphi$. Note that these terms
should be restored to get a full superconformal action, but that is not
the aim of this paper.  As the only fermionic terms in $\D_aL^{ij}$ are
proportional to $\varphi$, the first two terms on the RHS of
(\ref{bosonicconformal2}) will not contribute any new fermionic terms
when we write out the covariant derivatives. Instead we take a closer
look at the third term on the RHS of (\ref{bosonicconformal2}). Note that
$\varphi=0$ does not imply the vanishing of $\D_{\mu}\varphi^i$, see
(\ref{transformationslinear}). We obtain
\begin{eqnarray}
-\Linv L_{ij}\D_a\D^a L^{ij}&=&-\Linv L_{ij}e^{a \mu}\partial_{\mu}\D_a L^{ij}-\Linv L_{ij}e^{a \mu}\omega_{\mu a}{}^b\D_bL^{ij} \nonumber \\
          &&-2\Linv L_{ij}{\cal V}^{a}{}^{(i}{}_k\D_a L^{j)k}+8Lf_a{}^a \nonumber \\
&&+\Linv L_{ij}\opsi^{a i}\Bigl(-\frac{1}{2}\slashed{\mathcal{D}}L^{jk}\psi_{a k}+\frac{1}{4}\gamma^bE_b\psi_a^j+4L^{jk}\phi_{a k}\Bigr) \nonumber \\
&&+\frac{1}{6}L\opsi^a\gamma_{a}\chi\,. \label{bosonicconformalterm}
\end{eqnarray}
The last two lines will thus be added to ${\cal L}_{\rm ferm}$. Using
the first relation of (\ref{usefulexpr}), the first two terms in
(\ref{bosonicconformalterm}) combine into $-\Linv
L_{ij}e^{-1}\partial_{\mu}(e\D^{\mu} L^{ij})$ and there is a term
$-\frac{1}{2}\Linv L_{ij}\opsi^a\gamma_a\psi^b\D_bL^{ij}$ that needs to
be added to the fermionic Lagrangian. The bosonic Lagrangian thus
becomes
\begin{eqnarray}
e^{-1}\mathcal{L}_{\rm bos} &=& L^{-3}L_{ij}L_{kl}\D_aL^{kl}\D^aL^{ij}-\frac{1}{2}\Linv \D_a L_{ij} \D^a L^{ij}-\Linv L_{ij}e^{-1}\partial_{\mu}(e\D^{\mu} L^{ij}) \nonumber \\
&&-2\Linv L_{ij}{\cal V}^{a}{}^{(i}{}_k\D_a L^{j)k}+8Lf_a{}^a \nonumber \\
                         &&-\frac{1}{3}LD-\frac{1}{4}L^{-1}E_aE^a+E^{\mu}{\cal V}_{\mu ij}L^{ij}L^{-1}-\frac{1}{2}E^{\mu\nu}\partial_{\mu}L^{ij}L^k_j\partial_{\nu}L_{ik}L^{-3}. \label{bosonicconformal3}
\end{eqnarray}
Writing out the covariant derivatives $\D^aL_{ij}$,
dropping the fermionic terms (because they are proportional to $\varphi$
and hence vanish by gauge fixing) and terms proportional to\footnote{To fix the $K$-gauge, the condition $b_{\mu}=0$ will be imposed in the next section.} $b_{\mu}$
(for the same reason) we can write the bosonic Lagrangian as
\begin{eqnarray}
e^{-1}\mathcal{L}_{\rm bos} &=&8Lf_a{}^a+\frac{1}{2}L^{-1}\partial_aL^{ij}\partial^aL_{ij}-2\Linv L^k{}_i{\cal V}^{a}{}^{ij}\partial_a L_{kj}+2\Linv {\cal V}_a{}^{(l}{}_kL^{j)k}{\cal V}^a{}_{il}L^{i}{}_j \nonumber \\
                         &&-\frac{1}{3}LD-\frac{1}{4}L^{-1}E_aE^a+E^{\mu}{\cal V}_{\mu ij}L^{ij}L^{-1}-\frac{1}{2}E^{\mu\nu}\partial_{\mu}L^{ij}L^k_j\partial_{\nu}L_{ik}L^{-3}\,. \label{bosonicconformal4}
\end{eqnarray}
The fermionic action (with the $\varphi$ put to zero) looks at this point
as
\begin{eqnarray}
e^{-1}\mathcal{L}_{\rm ferm}&=&-\frac{1}{4}\Linv \opsi_{\rho}^i\gamma^{\mu\nu\rho}\psi_{\nu}^j L_{ij} E_{\mu}+\frac{1}{2}\Linv L^{ij}L_{kl}\opsi_{\rho}^k\gamma^{\mu\nu\rho}\psi_{\nu}^l {\cal V}_{\mu ij}-\frac{1}{3}L\opsi_{\mu}\gamma^{\mu}\chi \nonumber \\
             &&-\Linv L_{ij}\opsi_{\mu}^i\gamma^{\mu\nu}\Bigl(-\frac{1}{2}\slashed{\mathcal{D}}L^{jk}\psi_{\nu}{}_k+\frac{1}{4}\gamma^{\rho}E_{\rho}\psi_{\nu}{}^j+4L^{jk}\phi_{\nu}{}_k\Bigr) \nonumber \\
             &&+\frac{1}{6}L\opsi^a\gamma_{a}\chi-\frac{1}{2}\Linv L_{ij}\opsi^a\gamma_a\psi^b\D_bL^{ij}. \label{fermionicconformal2}
\end{eqnarray}
Note that this fermionic Lagrangian differs from the one in
(\ref{fermionicconformal1}) by the two last terms, and the $\gamma^\mu
\gamma ^\nu $ being replaced by $\gamma ^{\mu \nu }$, which originate
from the last two lines of (\ref{bosonicconformalterm}), and from $e^{a
\mu}\omega_{\mu a}{}^b$ in (\ref{usefulexpr}).

The bosonic terms in the expression of $f_a{}^a$ in (\ref{usefulexpr})
lead to
\begin{eqnarray}
e^{-1}\mathcal{L}_{\rm bos} &=&\frac{2}{5}L R+\frac{1}{2}L^{-1}\partial_aL^{ij}\partial^aL_{ij}-2\Linv L^k{}_i{\cal V}^{a}{}^{ij}\partial_a L_{kj} \nonumber \\
&&+2\Linv {\cal V}_a{}^{(l}{}_kL^{j)k}{\cal V}^a{}_{il}L^{i}{}_j-\frac{2}{15}L D-\frac{1}{4}L^{-1}E_aE^a+E^{\mu}{\cal V}_{\mu ij}L^{ij}L^{-1} \nonumber \\
&&-\frac{1}{2}E^{\mu\nu}\partial_{\mu}L^{ij}L^k_j\partial_{\nu}L_{ik}L^{-3}. \label{bosonicconformal5}
\end{eqnarray}
The fermionic terms of $f_a{}^a$ lead to terms modifying
$\mathcal{L}_{\rm ferm}$ to
\begin{eqnarray}
e^{-1}\mathcal{L}_{\rm ferm}&=& \frac{1}{2}\Linv L^{ij}L_{kl}\opsi_{\rho}^k\gamma^{\mu\nu\rho}\psi_{\nu}^l {\cal V}_{\mu ij}+\frac{1}{2}\Linv L_{ij}\opsi_{\mu}^i\gamma^{\mu\nu}\gamma^{\rho}D_{\rho}L^{jk}\psi_{\nu}{}_k \nonumber \\
             &&-L\opsi_{\mu}\gamma^\mu\left(2\gamma ^\nu\phi_{\nu}+\frac{1}{3}\chi\right)-\frac{4}{5}L\opsi^{\mu }\gamma^{\nu }\hat{D}_{[\nu }\psi_{\mu ]}+\frac{1}{5}L\opsi_b\gamma_c\psi_aT^-{}^{abc} \nonumber\\
&&-\frac{1}{2}\opsi^a\gamma_a\psi^b\partial _bL.
\label{finalfermionicconformal}
\end{eqnarray}
We used here the $\SU(2)$-covariant derivative,
\begin{equation}
  D_\mu  L^{ij}=\partial_{\mu}L^{ij}+2{\cal V}_{\mu}{}^{(i}{}_kL^{j)k},
 \label{DmuLij}
\end{equation}
where we already put $b_\mu =0$ in view of the gauge fixing of special
conformal transformations that we will adopt soon.
\subsection{Gauge fixing} \label{subsectionGaugefixing}

\subsubsection{Bosonic part}
To arrive at the super Poincar{\'e} group, the redundant symmetries of the
superconformal algebra need to be broken. The special conformal
transformations are fixed by the condition\footnote{We will keep the
notation $\hat{D}$ for the covariant derivative, but remember that from
now on it denotes $\hat{D}|_{\text{gauge fixed}}$.} $b_{\mu}=0$. The
dilatation gauge is fixed by $L=1$. The $\SU(2)$ symmetry cannot be
completely broken. A gauge choice $L^{ij}=\sqrt{\frac{1}{2}}\delta^{ij}$
still leaves a remaining $U(1)$ symmetry which will be gauged by the
auxiliary ${\cal V}_{\mu ij}\delta^{ij}$. For the bosonic part of the
gauge-fixed action, we use (\ref{LL}) to write
\begin{equation}
 2 {\cal V}_a{}^{(\ell }{}_kL^{j)k}{\cal V}^a{}_{i\ell }L^{i}{}_j={\cal V}'_a{}^{ij}\,{\cal V}'_{aij}\,.
 \label{berekeningbosVV}
\end{equation}
Here ${\cal V}'_{aij}$ is the traceless part of ${\cal V}_{aij}$:
\begin{equation}
  {\cal V}'_{aij}={\cal V}_{aij}-\ft12\delta _{ij}\delta ^{k\ell }{\cal V}_{ak\ell }.
 \label{Vtraceless}
\end{equation}
Applying the gauge fixing in (\ref{bosonicconformal5}), we obtain the
bosonic part of the gauge fixed action
\begin{eqnarray}
e^{-1}\mathcal{L}_{\rm bos} &=& \frac{2}{5}R+{\cal V}'_a{}^{ij}{\cal V}^{'a}{}_{ij} \nonumber \\
                         &&-\frac{2}{15}D-\frac{1}{4}E_aE^a+\frac{1}{\sqrt{2}}E^{\mu}{\cal V}_{\mu ij}\delta^{ij}.
\label{poincarebos1}
\end{eqnarray}

\subsubsection{Fermionic part}
We still need to fix the S-gauge. As mentioned in the previous section, this can be done by demanding $\varphi=0$. The fermionic part of the resulting action after gauge fixing is then
\begin{eqnarray}
e^{-1}\mathcal{L}_{\rm ferm}&=&-\frac{1}{4}{\cal V}_{\rho}{}^{kl}\Bigl(\delta_{ij}\delta_{kl}-\delta_{ik}\delta_{jl}+\epsilon_{kj}\epsilon_{li}\Bigr)\opsi_{\mu}^i\gamma^{\mu\nu\rho}\psi_{\nu}{}^j \nonumber \\
             &&-\frac{1}{2}\delta_{ij}\opsi_{\mu}^i g^{\rho[\mu}\gamma^{\nu]}{\cal V}_{\rho}{}^{j}{}_l\delta^{kl}\psi_{\nu}{}_k
             -\frac{1}{2}\opsi_{\mu}^ig^{\rho[\mu}\gamma^{\nu]}{\cal V}_{\rho}{}^{k}{}_i\psi_{\nu}{}_k\nonumber \\
             &&-\opsi_{\mu}\gamma^{\mu}\Bigl(2\gamma^{\nu}\phi_{\nu}+\frac{1}{3}\chi^i\Bigr)-\frac{4}{5}\opsi^{\mu }\gamma^{\nu }\hat{D}_{[\nu }\psi_{\mu ]}+\frac{1}{5}\opsi_b\gamma_c\psi_aT^-{}^{abc},
\end{eqnarray}
This can still be simplified using $\delta^{ij}\delta^{kl}-\delta^{il}\delta^{jk}=\varepsilon^{ik}\varepsilon^{jl}$ and the fact that $\delta_{ij}\opsi^i_{[\mu}\gamma^a\psi^j_{\nu]}=0$:
\begin{eqnarray}
e^{-1}\mathcal{L}_{\rm ferm}&=&-\frac{1}{2}{\cal V}_{\rho ij}\opsi_{\mu}^i\gamma^{\mu\nu\rho}\psi_{\nu}{}^j-\opsi_{\mu}\gamma^{\mu}\Bigl(2\gamma^{\nu}\phi_{\nu}+\frac{1}{3}\chi^i\Bigr)-\frac{4}{5}\opsi^{\mu }\gamma^{\nu }
\hat{D}_{[\nu }\psi_{\mu ]} \nonumber \\
&&+\frac{1}{5}\opsi_b\gamma_c\psi_aT^-{}^{abc}\,.
\end{eqnarray}
Next we use (\ref{usefulexpr}) to write
\begin{equation}
e^{-1}\mathcal{L}_{\rm ferm}=-\frac{1}{2}{\cal V}_{\rho ij}\opsi_{\mu}^i\gamma^{\mu\nu\rho}\psi_{\nu}{}^j-\frac{2}{5}\opsi_{\rho}\gamma^{\mu\nu\rho}\hat{D}_{\mu}\psi_{\nu}+\frac{1}{5}\opsi_b\gamma_c\psi_aT^-{}^{abc}-\frac{2}{15}\opsi_{\mu}\gamma^{\mu}\chi. \label{poincareferm1}
\end{equation}

\subsection{
Off-shell action} \label{subsectionPoincareWeyl2}

As can be seen from the Lagrangians (\ref{poincarebos1}) and
(\ref{poincareferm1}) the matter fields of the Weyl 1 multiplet, $D$,
$\chi^i$ and $T^-_{abc}$, have no kinetic terms\footnote{The same is true
for $E_a$ and ${\cal V}_{\mu}^{ij}$, but their field equations are
consistent.}. Also, the field equation for $D$ gives an inconsistency. As
mentioned in section \ref{subsectionWeyl2} this problem can be solved by
using the Weyl 2 multiplet instead. This can be done by plugging in the
expressions (\ref{tensorD}), (\ref{tensorchi}) and (\ref{tensorB}) into
the Lagrangians. We first write out the full expressions for $\chi^i$ and
D. From (\ref{tensorD}), (\ref{tensorchi}), (\ref{gamphiversion2}) and
(\ref{tensorB}) we find
\begin{eqnarray}
\chi^i&=&\frac{15}{4}\sigma^{-1}\gamma^{\mu}\bigl(D_{\mu}\psi^i-\frac{1}{48}\gamma\cdot\hat{F}\psi_{\mu}^i-\frac{1}{4}\hat{\slashed{D}}\sigma\psi_{\mu}^i\bigr) \nonumber \\
&&+\frac{3}{4}\gamma^{\mu\nu}\bigl(D_{\mu}\psi_{\nu}{}^i-\frac{1}{48}\sigma^{-1}\gamma\cdot\hat{F} \gamma_{\nu}\psi_{\mu}^i\bigr)-\frac{5}{32}\sigma^{-2}\gamma \cdot \hat{F}\psi^i,
\end{eqnarray}
and
\begin{eqnarray}
D&=&\frac{15}{4}\sigma^{-1}\Bigl[e^{-1}\partial_{\mu}(e\hat{D}^{\mu}\sigma)+\frac{1}{2}\opsi^a\gamma_a\psi^b\hat{D}_b\sigma-\frac{1}{5}\sigma R+\frac{1}{12}\sigma^{-1}\hat{F} \cdot \hat{F} \nonumber \\
&&+\frac{2}{5}\sigma\opsi^\mu \gamma^\nu \bigl(D_{[\nu }\psi_{\mu ]}-\frac{1}{48}\sigma^{-1}\gamma\cdot\hat{F}\gamma_{[\mu}\psi_{\nu]}\bigr)-\frac{1}{20}\opsi_b\gamma_c\psi_a\hat{F}^{-abc} \nonumber \\
&&-\opsi^{\mu}\bigl(D_{\mu}\psi-\frac{1}{48}\gamma\cdot\hat{F}\psi_{\mu}-\frac{1}{4}\hat{\slashed{D}}\sigma\psi_{\mu}\bigr)+\frac{1}{48}\sigma^{-1}\opsi\gamma\cdot \hat{F} \gamma^{\mu}\psi_{\mu} \nonumber \\
&&+\opsi\gamma^{\mu\nu}\bigl(D_{\mu}\psi_{\nu}-\frac{1}{48}\sigma^{-1}\gamma\cdot\hat{F} \gamma_{\nu}\psi_{\mu}\bigr)-\frac{1}{6}\sigma^{-2}\bar{\psi}\gamma \cdot \hat{F}\psi \nonumber \\
&&+4\sigma^{-1}\bar{\psi}\gamma^{\mu}\bigl(D_{\mu}\psi-\frac{1}{48}\gamma\cdot\hat{F}\psi_{\mu}-\frac{1}{4}\hat{\slashed{D}}\sigma\psi_{\mu}\bigr)\Bigr]\,, \label{finalD}
\end{eqnarray}
where the $D_{\mu}\psi_{\nu}^i$ and $D_{\mu}\psi^i$ are Lorentz and
$\rm{SU}(2)$ covariant derivatives, while $\hat{D}_{\mu}\sigma$ is a
supercovariant derivative:
\begin{eqnarray}
D_{\mu}\psi_{\nu}^i& \equiv \nabla_{\mu}\psi_{\nu}^i+{\cal V}_{\mu}{}^i{}_j\psi_{\nu}^j=&\bigl(\partial_{\mu}+\frac{1}{4}\omega_{\mu}{}^{ab}\gamma_{ab}\bigr)\psi_{\nu}^i
+{\cal V}_{\mu}{}^i{}_j\psi_{\nu}^j\,, \nonumber \\
D_{\mu}\psi^i& \equiv \nabla_{\mu}\psi^i+{\cal V}_{\mu}{}^i{}_j\psi^j=&\bigl(\partial_{\mu}+\frac{1}{4}\omega_{\mu}{}^{ab}\gamma_{ab}\bigr)\psi^i+{\cal V}_{\mu}{}^i{}_j\psi^j\,,\nonumber\\
\hat{D}_{\mu}\sigma&=&\partial_{\mu}\sigma-\opsi_{\mu}\psi. \label{SU2covder}
\end{eqnarray}
We also used (\ref{selfdualityrelations}) and the techniques of
(\ref{gammadotFrelation}).

The expressions for $\chi^i$ and $D$ can still be rewritten by using
several gamma matrix manipulations:
\begin{eqnarray}
\chi^i&=&\frac{15}{4}\sigma^{-1}\gamma^{\mu}D_{\mu}\psi^i+\frac{3}{4}\gamma^{\mu\nu}D_{\mu}\psi_{\nu}^i-\frac{15}{16}\sigma^{-1}\gamma^{\mu}\hat{\slashed{D}}\sigma\psi_{\mu}^i\nonumber \\
&&-\frac{3}{32}\sigma^{-1}\gamma^{\mu\rho\sigma\chi}\hat{F}_{\rho\sigma\chi}\psi_{\mu}^i-\frac{3}{16}\sigma^{-1}\gamma^{\sigma\chi}\hat{F}_{\rho\sigma\chi}\psi^{\rho}{}^i-\frac{5}{32}\sigma^{-2}\gamma \cdot \hat{F}\psi^i
\end{eqnarray}
and
\begin{eqnarray}
D&=&\frac{15}{4}\sigma^{-1}\Bigl[e^{-1}\partial_{\mu}(e\hat{D}^{\mu}\sigma)+\frac{1}{2}\opsi^a\gamma_a\psi^b\hat{D}_b\sigma-\frac{1}{5}\sigma R+\frac{1}{12}\sigma^{-1}\hat{F} \cdot \hat{F} \nonumber \\
&&+\frac{2}{5}\sigma\opsi^\mu \gamma^\nu D_{[\nu }\psi_{\mu ]}-\frac{1}{40}\opsi^{\rho}\gamma^{\mu\sigma\chi}\hat{F}_{\rho\sigma\chi}\psi_{\mu}-\opsi^{\mu}D_{\mu}\psi+\frac{1}{40}\opsi^{\mu}\gamma\cdot\hat{F}\psi_{\mu} \nonumber \\
&&+\opsi\gamma^{\mu\nu}D_{\mu}\psi_{\nu}-\frac{1}{6}\sigma^{-2}\bar{\psi}\gamma \cdot \hat{F}\psi-\frac{1}{8}\sigma^{-1}\bar{\psi}\gamma^{\mu\nu\rho\sigma}\hat{F}_{\nu\rho\sigma}\psi_{\mu}-\frac{1}{8}\sigma^{-1}\bar{\psi}\gamma^{\rho\sigma}\hat{F}_{\nu\rho\sigma}\psi^{\nu} \nonumber \\
&&+4\sigma^{-1}\bar{\psi}\slashed{D}\psi-\sigma^{-1}\bar{\psi}\gamma^{\mu}\hat{\slashed{D}}\sigma\psi_{\mu}\Bigr]\,,
\label{finalexprD}
\end{eqnarray}
where we also used
\begin{equation}
\gamma_{\rho}\F^-{}^{\mu\nu\rho}=\frac{1}{2}\gamma_{\rho}\bigl(\F^{\mu\nu\rho}+\frac{1}{6}\epsilon^{\mu\nu\rho\sigma\lambda\chi}\F_{\sigma\lambda\chi}\bigr)
=\frac{1}{2}\bigl(\gamma_{\rho}\F^{\mu\nu\rho}+\frac{1}{6}\gamma^{\mu\nu\sigma\lambda\chi}\gamma_*\F_{\sigma\lambda\chi}\bigr).
\end{equation}
This equation can be proven using the duality relation
(\ref{dualityrelation}).

Filling in the expressions for $D$ and $\chi^i$ into (\ref{poincarebos1}) and (\ref{poincareferm1}) and distributing the respective terms over the bosonic and fermionic Lagrangians we get
\begin{eqnarray}
e^{-1}\mathcal{L}_{\rm bos} &=&\frac{1}{2}R+{\cal V}'_a{}^{ij}{\cal V}^{'a}{}_{ij}-\frac{1}{4}E_aE^a+\frac{1}{\sqrt{2}}E^{\mu}{\cal V}_{\mu ij}\delta^{ij} \nonumber \\
                         &&-\frac{1}{2}\sigma^{-2}\partial_a\sigma\partial^a\sigma-\frac{3}{8}\sigma^{-2}\partial_{[\mu}B_{\nu\rho]}\partial^{\mu}B^{\nu\rho}\,,
                         \label{poincarebos2}
\end{eqnarray}
where we used the definition of $\hat{F}(B)$ in (\ref{weyl2FB}). We then
perform gamma matrix manipulations and write out the covariant
derivatives (\ref{SU2covder}). After dropping a total derivative, the
fermionic Lagrangian becomes
\begin{eqnarray}
e^{-1}\mathcal{L}_{\rm ferm}&=&-\frac{1}{2}\opsi_{\rho}\gamma^{\mu\nu\rho}\nabla_{\mu}\psi_{\nu}-2\sigma^{-2}\bar{\psi}\slashed{D}\psi+\sigma^{-2}\opsi\gamma^{\nu}\gamma^{\mu}\psi_{\nu}\partial_{\mu}\sigma \nonumber \\
&&+\frac{1}{16}\sigma^{-1}\opsi_{\mu}\gamma^{\mu\nu\rho\sigma\chi}\psi_{\nu}\partial_{\rho}B_{\sigma\chi}-\frac{3}{8}\sigma^{-1}\opsi^{\mu}\gamma^{\nu}\psi^{\rho}\partial_{[\mu}B_{\nu\rho]} \nonumber \\
&&+\frac{1}{4}\sigma^{-2}\opsi_{\mu}\gamma^{\mu\rho\sigma\chi}\psi\partial_{\rho}B_{\sigma\chi}-\frac{3}{4}\sigma^{-2}\opsi^{\mu}\gamma^{\nu\rho}\psi\partial_{[\mu}B_{\nu\rho]} \nonumber \\
&&+\frac{1}{4}\sigma^{-3}\bar{\psi}\gamma^{\mu\nu\rho}\psi\partial_{\mu}B_{\nu\rho}+\Bigl(\text{4-fermion terms}\Bigr), \label{poincareferm2}
\end{eqnarray}
where
\begin{eqnarray}
\Bigl(\text{4-fermion terms}\Bigr)&=&\frac{1}{32}\opsi_{\mu}\gamma^{\mu\nu\rho\sigma\chi}\psi_{\nu}\opsi_{\rho}\gamma_{\sigma}\psi_{\chi}-\frac{3}{32}\opsi_{[\mu}\gamma_{\nu}\psi_{\rho]}\opsi^{\mu}\gamma^{\nu}\psi^{\rho} \nonumber \\
&&-\frac{1}{4}\sigma^{-1}\opsi^{\mu}\gamma_{\mu}\psi^{\nu}\opsi^{\rho}\gamma_{\rho\nu}\psi+\frac{1}{8}\sigma^{-1}\opsi_{\mu}\gamma^{\nu}\psi_{\sigma}\opsi_{\nu}\gamma^{\mu\sigma}\psi \nonumber \\
&&+\frac{1}{16}\sigma^{-1}\opsi_{\mu}\gamma^{\mu\nu\rho\sigma\chi}\psi_{\nu}\opsi_{\rho}\gamma_{\sigma\chi}\psi-\frac{3}{8}\sigma^{-1}\opsi_{[\mu}\gamma_{\nu\rho]}\psi\opsi^{\mu}\gamma^{\nu}\psi^{\rho} \nonumber \\
&&+\frac{1}{8}\sigma^{-1}\opsi_{\mu}\gamma^{\mu\rho\sigma\chi}\psi\opsi_{\rho}\gamma_{\sigma}\psi_{\chi} \nonumber \\
&&+\frac{1}{4}\sigma^{-2}\opsi_{\mu}\gamma^{\mu\rho\sigma\chi}\psi\opsi_{\rho}\gamma_{\sigma\chi}\psi-\frac{3}{8}\sigma^{-2}\opsi_{[\mu}\gamma_{\nu\rho]}\psi\opsi^{\mu}\gamma^{\nu\rho}\psi \nonumber \\
&&-\frac{1}{2}\sigma^{-2}\opsi\gamma^{\mu\nu}\psi_{\mu}\opsi_{\nu}\psi-\frac{1}{2}\sigma^{-2}\bar{\psi}\psi^{\nu}\opsi_{\nu}\psi+\frac{1}{8}\sigma^{-2}\opsi\gamma^{\rho\sigma\chi}\psi\opsi_{\rho}\gamma_{\sigma}\psi_{\chi} \nonumber \\
&&+\frac{1}{4}\sigma^{-3}\opsi\gamma^{\rho\sigma\chi}\psi\opsi_{\rho}\gamma_{\sigma\chi}\psi.
\end{eqnarray}
\section{Conclusions} \label{ss:conclusions}

In this paper we constructed the full (bosonic and fermionic) off-shell
action of minimal Poincar{\'e} supergravity in six dimensions. We obtained
this action by using the methods of superconformal tensor calculus as
suggested in \cite{Bergshoeff:1985mz}. We constructed a superconformal
action by coupling a linear compensator multiplet (which is an off-shell
multiplet) to the Weyl 1 multiplet in a density formula. By fixing the
redundant symmetries ($D$, $K$ and $S$) we obtained the Poincar{\'e} action.
However, this action contained no kinetic terms for the matter fields of
the Weyl 1 multiplet and even led to an inconsistent field equation for
the $D$ field. This forced us to consider the matter fields of the Weyl 1
multiplet as functions of those of the Weyl 2 multiplet. Expressing the
action in terms of fields of the Weyl 2 multiplet led to an action that
contains kinetic terms for the matter fields and has consistent field
equations.

In this sense the six dimensional case differs from the four and five
dimensional one. In four and five dimensions
\cite{deWit:1979ug,Kugo:2000hn,Bergshoeff:2001hc,Fujita:2001kv} the Weyl
1 multiplet can be used as an independent multiplet to construct a pure
Poincar{\'e} action, not leading to inconsistencies and having proper kinetic
terms for the matter fields.

Another fact worth mentioning is the following. We constructed the
superconformal action by using a density formula for the product of a
linear and a vector multiplet. This formula could be used to construct an
action for the linear compensator provided we were able to define a
vector multiplet from the components of the linear multiplet. One way to
do this is given in appendix \ref{app:vectorfromlinear} in which a linear
multiplet is coupled to Weyl 1. Factors of $L$ are used to compensate for
the difference in Weyl weights between the linear and the vector
multiplet components. After gauge fixing, the Weyl 1 matter fields were
solved in terms of Weyl 2 matter fields. Another possibility to construct
a vector multiplet from the components of a linear one is to couple
directly to the Weyl 2 multiplet and use factors of $\sigma$ to account
for the different Weyl weights. Both possibilities can be generalized by
considering a combination $L^{\alpha}\sigma^{2(1-\alpha)}$ as
compensator. We expect, of course, this free parameter $\alpha$ to
disappear in the gauge fixing procedure, hence leading to the same
off-shell Poincar{\'e} action for all $\alpha$.

By performing the gauge fixing, the $\rm{SU}(2)$ R-symmetry of the
superconformal group is broken to a $\rm{U}(1)$. In the final Poincar{\'e}
action this $\rm{U}(1)$ is gauged by the auxiliary ${\cal
V}_{\mu}^{ij}\delta_{ij}$. As mentioned in the introduction, it would be
interesting to add a vector multiplet to the action. This physical
vector multiplet can then be coupled to the $\rm{U}(1)$ symmetry by
adding a coupling of the form (\ref{densityformula})
\cite{Bergshoeff:1985mz}, thus obtaining the dual off-shell Salam-Sezgin
model. Adding $R^2$-terms to the action in a supersymmetric way (as
described in \cite{Bergshoeff:1986vy, Bergshoeff:1986wc}), would be
interesting to study the influence of these terms on solutions.

\section*{Acknowledgements}

We are grateful to Eric Bergshoeff and Ergin Sezgin for useful
discussions. F.C. also wants to thank the Centre for Theoretical Physics
of the University of Groningen for the hospitality during his visit to
Groningen. This work is supported in part by the FWO - Vlaanderen,
Project No. G.0235.05, and in part by the Federal Office for Scientific,
Technical and Cultural Affairs through the ``Interuniversity Attraction
Poles Programme -- Belgian Science Policy'' P6/11-P.
\newpage
\appendix

\section{Notation and conventions} \label{ss:conventions}

Note that some conventions differ from the ones used in
\cite{Bergshoeff:1985mz}. When this is the case we will explicitly
mention it.

We use the $(-+\ldots +)$ metric (as opposed to the Pauli metric
$(++\ldots +)$ in \cite{Bergshoeff:1985mz}) and the $8\times8$ Dirac
matrices $\gamma_a$ $(a=0,\ldots,5)$ are defined by the property
\begin{equation}
\gamma_a\gamma_b+\gamma_b\gamma_a=2\eta_{ab},
\end{equation}
and are Hermitian. A complete set of $8\times8$ matrices
 is given by
\begin{equation}
O_I=\{\textbf{1},\gamma^{(1)},\gamma^{(2)},\gamma^{(3)},\gamma^{(4)},\gamma^{(5)},\gamma^{(6)}\}, \label{Cliffordbasis}
\end{equation}
where we have used the following notation
\begin{equation}
\gamma^{(n)}=\gamma^{a_1\ldots a_n}=\gamma^{[a_1}\gamma^{a_2}\ldots\gamma^{a_n]}=\frac{1}{n!}\sum_p(-1)^p\gamma^{a_1}\ldots\gamma^{a_n},
\label{higherrankgammas}
\end{equation}
where $\sum_p$ means summation over all permutations. The matrix $\gamma_*$ is defined by
\begin{equation}
\gamma_*=\gamma_0\gamma_1\gamma_2\gamma_3\gamma_4\gamma_5.
\end{equation}
This definition ensures that it squares to one (note that this differs
from \cite{Bergshoeff:1985mz}). It can be used to define left and right
handed spinors:
\begin{equation}
  P_L=\ft12(1 +\gamma _*)\,,\qquad P_R=\ft12(1 -\gamma _*).
 \label{gamma*}
\end{equation}
The spinors are symplectic. We define the supersymmetries as left-handed,
i.e.
\begin{equation}
  \epsilon ^i = P_L\epsilon ^i\,,\qquad \bar \epsilon ^i = \bar \epsilon ^i P_R.
 \label{chiralD6eps}
\end{equation}
For the other fermion fields we use
\begin{eqnarray}
  &\psi _\mu ^i=P_L\psi _\mu ^i\,, \qquad \chi ^i=P_L\chi ^i\,, \qquad \Omega ^i=P_L\Omega ^i,& \nonumber \\
  &\phi _\mu ^i=P_R\phi _\mu ^i\,,\qquad \psi ^i=P_R\psi ^i\,,\qquad \varphi ^i =P_R\varphi ^i\,.&
 \label{LRfermions}
\end{eqnarray}

Indices $i$, $j$ are raised and lowered by the $\varepsilon _{ij}$ in
$NW-SE$ direction, i.e.
\begin{equation}
  \lambda _i = \lambda ^j\varepsilon _{ji}\,,\qquad \lambda ^i =\varepsilon ^{ij}\lambda _j\,.
 \label{raiselowersympl}
\end{equation}
With the raising and lowering conventions of $\SU(2)$ indices as in
(\ref{raiselowersympl}), one has for any two objects $A,B$:
$A^iB_i=-A_iB^i$. When $\rm{SU}(2)$ indices are omitted, a $NW-SE$ contraction is understood, e.g.
\begin{equation}
\bar{\lambda}\gamma^{(n)}\psi=\bar{\lambda}^i\gamma^{(n)}\psi_i.
\end{equation}
Changing the order of spinors in a bilinear leads to the following signs:
\begin{equation}
\bar{\lambda}^i\gamma^{(n)}\psi^j=t_n\bar{\psi}^j\gamma^{(n)}\lambda^i, \qquad
 t_n=\begin{cases}+\text{: } n=1,2,5,6 \\ -\text{: } n=0,3,4\end{cases}
\end{equation}
An additional sign is needed if the $\rm{SU}(2)$ indices are contracted,
e.g. $\bar{\lambda}\gamma^a\psi=-\bar{\psi}\gamma^a\lambda$ but
$\bar{\lambda}^i\gamma^a\psi^j=\bar{\psi}^j\gamma^a\lambda^i$.

For any antisymmetric tensor $A^{ij}$ one can write $A^{ij}=
\ft12\varepsilon ^{ij}A_k{}^k$.
Under charge conjugation (which is equal to complex conjugation on scalar
quantities)
\begin{equation}
  \lambda _i= (\lambda ^i)^C\,,\qquad \lambda ^i=-(\lambda _i)^C\,.
 \label{lambdaiCC}
\end{equation}

The totally antisymmetric rank 6 tensor is denoted by $\varepsilon_{abcdef}$ or $\varepsilon^{abcdef}$ with
\begin{equation}
\varepsilon_{012345}=1, \hspace{2cm}  \varepsilon^{012345}=-1.
\end{equation}
It satisfies
\begin{eqnarray}
\varepsilon_{a_1 a_2 \ldots a_n b_1 \ldots b_{6-n}}\varepsilon^{a_1 a_2 \ldots a_n c_1 \ldots c_{6-n}}&=&-(6-n)!n!\delta_{b_1 \ldots b_{6-n}}^{[c_1 \ldots c_{6-n}]}. \label{epsilonrelations}
\end{eqnarray}
Throughout this paper the dual $\tilde{T}_{abc}$ of a tensor $T_{abc}$ is defined by
\begin{equation}
\tilde{T}_{abc}=-\frac{1}{6}\varepsilon_{abcdef}T^{def},
\label{deftilde6}
\end{equation}
which means that $\tilde{\tilde{T}}=T$. In \cite{Bergshoeff:1985mz} there
is an $\rmi$ factor in (\ref{deftilde6}), which  is related to the use of
the Pauli metric. Positive and negative dual parts are defined by
\begin{equation}
T^{\pm}_{abc}=\frac{1}{2}\bigl(T_{abc}\pm\tilde{T}_{abc}\bigr).
\end{equation}
Following duality relation will also prove to be useful
\begin{equation}
\gamma^{a_1 a_2 \ldots a_r}\gamma_*=-\frac{1}{(6-r)!}\varepsilon^{a_r a_{r-1} \ldots a_1 b_1 b_2 \ldots b_{D-r}}\gamma_{b_1 b_2 \ldots b_{D-r}}.
\label{dualityrelation}
\end{equation}
It implies that
\begin{equation}
  \gamma _{abc}\gamma _*= -\tilde \gamma ^{abc}, \qquad \gamma_{abc}P_L= \gamma _{abc}^-.
 \label{tildegamma}
\end{equation}
Finally we remark that in $D=6$ the product of two tensors with opposite
duality is non-zero but the product of two tensors of the same duality
vanishes:
\begin{eqnarray}
T^+_{abc}T^-{}^{abc}&=&\frac{1}{2}T_{abc}T^{abc}\,, \nonumber \\
T^{+}_{abc}T^{+}{}^{abc}&=&T^{-}_{abc}T^{-}{}^{abc}=0\,. \label{selfdualityrelations}
\end{eqnarray}
This can be combined with (\ref{tildegamma}) to write e.g.
\begin{eqnarray}
&& \gamma \cdot F \psi_{\mu}^i= \gamma \cdot F P_L \psi_{\mu}^i=\gamma^-\cdot F\psi_{\mu}^i= \gamma^-\cdot F^+\psi_{\mu}^i= \gamma\cdot F^+\psi_{\mu}^i\,,\nonumber\\
&& \gamma \cdot F \psi^i= \gamma \cdot F P_R \psi^i=\gamma^+\cdot F\psi^i= \gamma^+\cdot F^-\psi^i= \gamma\cdot F^-\psi^i.
 \label{gammadotFrelation}
\end{eqnarray}

\section{The superconformal algebra in six dimensions} \label{ss:6dsca}
The superalgebra that we gauge is $\OSp(8^*|2)$.  The conformal algebra
$\SO(6,2)= \SO^*(8)$ is
\begin{eqnarray}
&&[M_{\mu \nu } , M_{\rho \sigma }]\,=\,4\eta_{[\mu [\rho } M_{\sigma ]\nu ]}=\eta_{\mu \rho } M_{\sigma \nu }-\eta_{\nu \rho } M_{\sigma \mu }
-\eta_{\mu \sigma  } M_{\rho \nu }+\eta_{\nu\sigma } M_{\rho  \mu }\,, \nonumber\\
&& [P_\mu  , M_{\nu \rho } ] \,=\,2\eta_{\mu [\nu } P_{\rho ]}\,, \qquad 
 [K_\mu  , M_{\nu \rho } ] \,=\, 2\eta_{\mu [\nu }K_{\rho ]}\,, \nonumber\\
&& [P_\mu  , K_\nu  ]\,=\, 2 (\eta_{\mu \nu } D +  M_{\mu \nu}) \,,
\nonumber\\
&& [D , P_\mu   ]\,=\, P_\mu  \,,\qquad \quad \quad
\ [D , K_\mu   ]\,=\, -K_\mu  \,.     \label{confalg}
\end{eqnarray}
The $\SU(2)$ algebra,
can be written as
\begin{equation}
  [U_i{}^j, U_k{}^\ell ] \,=\, \delta_i{}^\ell  U_k{}^j - \delta_k{}^j
U_i{}^\ell\,.
 \label{SU2algebraij}
\end{equation}
The fermionic generators are symplectic Majorana-Weyl spinors, with the
convention
\begin{equation}
  Q^i= P_R Q^i=-\gamma _*Q^i\,,\qquad S^i=P_LS^i=\gamma _*S^i\,.
 \label{chiralQSD6}
\end{equation}
The commutators between bosonic and fermionic generators are
\begin{equation}
\begin{array}{ll}
[M_{ab} , Q_\alpha^i ] \,=\,- \ft12 ( \gamma_{ab} Q^i)_\alpha\,,\qquad
& [M_{ab} , S_\alpha^i ] \,=\,- \ft12 ( \gamma_{ab} S^i)_\alpha\,,
\\{}
[D,Q_\alpha{}^i] \,=\,\ft12 Q_\alpha{}^i\,,\qquad & [D,S_\alpha{}^i]=-
\ft12 S_\alpha^i\,,
\\{}
[U_i{}^j, Q_\alpha{}^k] \,=\,\delta_i{}^k Q_\alpha^j - \ft 12
\delta_i{}^j Q_\alpha{}^k\,,\quad & {}[U_i{}^j, S_\alpha{}^k]\,=\,
\delta_i{}^k S_\alpha^j -
\ft 12 \delta_i{}^j S_\alpha{}^k\,,
\\{}
[U_i{}^j, Q_{\alpha k}] \,=\,-\delta_k{}^j Q_{\alpha i} + \ft 12
\delta_i{}^j Q_{\alpha k}\,,\quad & {}[U_i{}^j, S_{\alpha k}]\,=\,
\delta_k{}^j S_{\alpha i} -
\ft 12 \delta_i{}^j S_{\alpha k}\,,
\\ {}
[K_a, Q_{\alpha}^i] \,=\,-(\gamma_{a}
S^i)_\alpha\,, \qquad & [P_a, S_\alpha{}^i] \,=\,-( \gamma_{a}
Q^i)_\alpha\,.\end{array}
 \label{ConfCommutators}
\end{equation}

The anticommutation relations between the fermionic generators are (with
the convention that $Q_\alpha $ are the components of the spinors $Q$,
and $Q^\alpha $ those of $\bar Q=Q^TC$):
\begin{eqnarray}
 & & \{ Q_{i\alpha }, Q^{j\beta }\}\ = \
 -\ft 12\delta _i{}^j (\gamma ^a)_\alpha{}^\beta  P_a \, ,\qquad
\{ S_{i\alpha }, S^{j\beta } \}\ = \
-\ft 12 \delta _i{}^j (\gamma ^a)_\alpha{}^\beta  K_a \, , \nonumber\\
 &&\{ Q_{i\alpha }, S^{j\beta }\} \ = \ \ft 12 \left(\delta _i{}^j
\delta _\alpha {}^\beta  D +
     \ft12 \delta _i{}^j (\gamma ^{ab}) _\alpha {}^\beta  M_{ab}
      + 4\delta _\alpha {}^\beta  U_i{}^j \right)         \,.
 \label{anticommN2}
\end{eqnarray}
For readability of the formulas, we omitted in the right-hand side $P_L$
or $P_R$ projection matrices, which follow from the chirality properties
of the generators in the left-hand side.

\section{Construction of a vector multiplet from the components of a linear multiplet} \label{app:vectorfromlinear}

The linear multiplet was introduced in section \ref{sec:linear} with the
fields $L^{ij}$, $\varphi^i$ and $E_a$. We define
\begin{equation}
L=\bigl(L_{ij}L^{ij}\bigr)^{1/2},
\end{equation}
which implies
\begin{equation}
L_{ij}L^{jk}=\frac{1}{2}\delta_i^kL^2,\qquad L^{i}{}_jL^{jk}= \ft12\varepsilon ^{ik}L^2.
\label{LL}
\end{equation}
The following formula for the second derivative of $L$ (with any type of
covariant derivative $\D$) is also useful:
\begin{equation}
\D_a\D^aL=L^{-1}L^{ij}\D_a\D^aL_{ij}+L^{-1}\D_aL^{ij}\D^aL_{ij}-L^{-3}L^{ij}\D_aL_{ij}L^{kl}\D^aL_{kl}.
\label{dadaL}
\end{equation}

The $\cn=(1,0)$, $D=6$ abelian vector multiplet consists of a real vector
field $W_{\mu}$, an $\rm{SU}(2)$ Majorana spinor $\Omega^i$ of positive
chirality and a triplet of auxiliary scalar fields
$Y^{ij}=\bigl(Y^{ij}\bigr)^*$. The full nonlinear $Q$- and
$S$-transformation rules are given by:
\begin{eqnarray}
\delta W_{\mu}&=&-\bar{\epsilon}\gamma_{\mu}\Omega ,\nonumber \\
\delta \Omega^i&=&\frac{1}{8}\gamma\cdot\hat{F}(W)\epsilon^i-\frac{1}{2}Y^{ij}\epsilon_j, \nonumber \\
\delta Y^{ij}&=&-\bar{\epsilon}^{(i}\slashed{\mathcal{D}}\Omega^{j)}+2\bar{\eta}^{(i}\Omega^{j)}, \nonumber \\
\hat{F}_{\mu\nu}(W)&=&F_{\mu\nu}(W)+2\bar{\psi}_{[\mu}\gamma_{\nu]}\Omega, \nonumber \\
\D_{\mu}\Omega^i&=&\partial_{\mu}\Omega^i-\frac{3}{2}b_{\mu}\Omega^i+\frac{1}{4}\omega_{\mu}{}^{ab}\gamma_{ab}\Omega^i-\frac{1}{2}V_{\mu}{}^i{}_j\Omega^j \nonumber \\
&&-\frac{1}{8}\gamma\cdot\hat{F}(W)\psi_{\mu}^i+\frac{1}{2}Y^{ij}\psi_{\mu j}. \label{AbelianVM}
\end{eqnarray}
In \cite{Bergshoeff:1985mz} one constructs a vector multiplet as
multiplet of field equations for the fields of the linear multiplet. We
summarize the results:
\begin{eqnarray}
\Omega^i&=&\Linv\slashed{\mathcal{D}}\varphi^i-L^{-3}\bigl(\slashed{\mathcal{D}}L^{ij}\bigr)L_{jk}\varphi^k+\frac{1}{2}L^{-3}\gamma^aE_aL^{ij}\varphi_j+\frac{2}{3}\Linv L^{ij}\chi_j \nonumber \\
&&+\frac{1}{12}\Linv \gamma \cdot T^-\varphi^i+\frac{1}{2}L^{-5}L^{ij}\gamma_a\varphi_jL^{kl}\bar{\varphi}_k\gamma_a\varphi_l, \nonumber \\
Y^{ij}&=&-\Linv\D_a\D^aL^{ij}+L^{-3}L_{kl}\D^aL^{k(i}\D_aL^{j)l}+\frac{1}{4}L^{-3}E_aE^aL^{ij} \nonumber \\
&&-L^{-3}E^aL^{k(i}\D_aL^{j)}{}_k-\frac{1}{3}\Linv L^{ij}D+\frac{1}{6}\Linv\bar{\chi}^{(i}\varphi^{j)} \nonumber \\
&&-\frac{4}{3}L^{-3}L^{k(i}L^{j)l}\bar{\chi}_k\varphi_l+\frac{1}{4}L^{-3}L^{ij}\bar{\varphi}^k\slashed{\D}\varphi_k+2L^{-3}L^{k(i}\bar{\varphi}_k\slashed{\D}\varphi^{j)} \nonumber \\
&&-L^{-3}\D_aL^{k(i}\varphi^{j)}\gamma^a\varphi_k-3L^{-5}L^{pq}L^{k(i}\D_aL^{j)}{}_k\bar{\varphi}_p\gamma^a\varphi_q \nonumber \\
&&-\frac{1}{12}L^{-3}L^{ij}\bar{\varphi}^k\gamma\cdot T^-\varphi_k+\frac{1}{4}L^{-3}\bar{\varphi}^{(i}\gamma^aE_a\varphi^{j)} \nonumber \\
&&+\frac{3}{2}L^{-5}L^{k(i}L^{j)l}\bar{\varphi}_k\gamma^a\varphi_lE_a-\frac{1}{2}L^{-5}\bar{\varphi}^{(i}\gamma_a\varphi^{j)}L^{kl}\bar{\varphi}_k\gamma^a\varphi_l \nonumber \\
&&+\frac{5}{4}L^{-7}L^{ij}L^{kl}\bar{\varphi}_k\gamma_a\varphi_lL^{mn}\bar{\varphi}_m\gamma^a\varphi_n, \nonumber \\
\hat{F}_{ab}(W)&=&2\Linv L^{ij}\hat{R}_{abij}(V)-2\D_{[a}\bigl(\Linv E_{b]}\bigr)-2L^{-3}L^l{}_k\D_{[a}L^{kp}\D_{b]}L_{lp} \nonumber \\
&&+\Linv \bar{\hat{R}}_{ab}{}^k(Q)\varphi_k-2\D_{[a}\bigl(L^{-3}L^{ij}\bar{\varphi}_i\gamma_{b]}\varphi_j\bigr). \label{vectorlinear}
\end{eqnarray}
\newpage

\end{document}